\documentclass[aps,prd,twocolumn,superscriptaddress]{revtex4-1}

\usepackage[english]{babel}
\usepackage{amsmath, amsfonts, amsthm, amsbsy, amsopn, amsxtra, amscd, amstext}
\usepackage{longtable}
\usepackage{bbm}
\usepackage{enumitem}
\usepackage{graphicx}
\usepackage{xcolor}
\usepackage{wrapfig}
\usepackage{soul}
\usepackage{physics}
\usepackage{amssymb}
\usepackage[hidelinks]{hyperref}

\newcommand\E[1]{{\left\langle #1 \right \rangle}}
\newcommand\SkipFisher[1]{}

\newcommand*{\dlum}{d_{\mathrm{L}}}

\newcommand\mc{{\cal M}_c}
\newcommand\HideMe[1]{}

\newcommand{\AffiliationCCRG}{
  Center for Computational Relativity and Gravitation, 
  Rochester Institute of Technology, 
  Rochester, New York 14623, USA 
}

\newcommand{\AffiliationUWM}{
  Department of Physics, 
  University of Wisconsin--Milwaukee, 
  Milwaukee, WI 53201, USA 
}
\newcommand{\AffiliationGSFC}{
  NASA Postdoctoral Program, 
  Astrophysics Science Division, 
  NASA Goddard Space Flight Center, 
  Greenbelt, MD 20771, USA 
}

\begin{document}
\title{
    Compressed Parametric and Non-Parametric Approximations to the Gravitational Wave Likelihood
}
\author{V. Delfavero}
\email[]{msd8070@rit.edu}
\affiliation{\AffiliationGSFC}

\author{R. O'Shaughnessy}
\affiliation{\AffiliationCCRG}

\author{D. Wysocki}
\affiliation{\AffiliationUWM}

\author{A. Yelikar}
\affiliation{\AffiliationCCRG}

\date{\today}

\clearpage{}\begin{abstract}
Gravitational-wave observations of quasicircular compact binary mergers imply complicated posterior measurements of their parameters. Though Gaussian approximations to the pertinent likelihoods have decades of history in the field, the relative generality and practical utility of these approximations hasn't been appreciated, given focus on careful, comprehensive generic Bayesian parameter inference. Building on our previous work in three dimensions, we demonstrate by example that bounded multivariate normal likelihood approximations are a sufficiently accurate representation of the full likelihood of observed gravitational-wave sources. Fits for each event are published in the Gravitational-Wave Transient Catalogs at \url{https://gitlab.com/xevra/nal-data}, along with a code release at \url{https://gitlab.com/xevra/gwalk}. We argue our approximations are more than accurate enough for population inference and introduce much smaller errors than waveform model systematics. To demonstrate the utility of these approximations as parametric models for the likelihood of individual gravitational-wave sources, we show examples of their application to modeling the population of observed gravitational-wave sources as well as low-latency parameter inference.
\end{abstract}
\clearpage{}

\maketitle

\begin{section}{Introduction}
\label{sec:intro}
The first gravitational-wave observation was made by the Laser Interferometer
    Gravitational-wave Observatory (LIGO) in 2015
    \cite{GW150914-detection,GW150914-astro,GW150914-num}.
To date, the LIGO Scientific Collaboration, together with VIRGO and KAGRA,
    have reported the detection of 90 gravitational-wave signals from
    merging compact binaries in the Gravitational-Wave Transient Catalogs
    \cite{GWTC-1, GWTC-2, GWTC-2p1, GWTC-3, O3-Detector, Virgo, Kagra}.
Many groups have developed non-LVK catalogs, and while important, they are not
    the focus of our study
    \cite{IAS, 3-OGC, Boyle_2019_SxS, Jani_2016_GA_Tech, Jani_2016_GA_Tech, BAM2008}.
These merging compact binaries 
    include Binary Black Hole (BBH), Binary Neutron Star (BNS), and
    Neutron-Star-Black-Hole (NSBH) mergers.
As each observation is merely a strain signal identified in each interferometer,
    extracting information from these signals requires a 
    Bayesian statistical inverse technique:
        gravitational-wave parameter estimation.

Many groups have developed methodology to evaluate the agreement
    of a gravitational-wave signal to relativistic waveform models
    in a space of these parameters
    \cite{Pankow2015,Veitch2015LALInference,2019PASP..131b4503B,RIFT,Bilby2019,IAS}.
There is additional consideration for events which may have electromagnetic
    counterparts, and a low latency parameter estimation pipeline
    must identify and localize such detections
    \cite{Aasi_2014,Singer_2016, PhysRevLett.127.241103,
        Evans_2012,LIGOLowLatencyUserGuide,
        Finstad_2020, Morisaki_2020}.
These groups strive to yield fast and accurate parameter estimates
    for the parameters of each event.
The parameters associated with a given binary can be split into \textit{extrinsic}
    parameters which depend on the orientation of a source relative
    to the observer, and the \textit{intrinsic} parameters which do not.
The astrophysical parameters of interest to a study of the 
    underlying population of merging binaries include the intrinsic parameters
    and distance ($\dlum$).
The intrinsic parameters of a binary black hole system (BBH)
    include the mass of each companion ($m_i$)
    as well as the dimensionless spin ($\mathbf{\chi_i}$).
For binary neutron star mergers, the tidal deformability of each
    companion ($\Lambda_i$) is also estimated.
Collectively, these are the astrophysical parameters ($x$) of
    the compact binary merger.
These describe a system without reference to the observer,
    and fully characterize the astrophysical properties of
    the underlying population of merging binaries.

Various methods are used to estimate various marginal likelihood functions for
    the astrophysical parameters of gravitational-wave events,
    after a standard Bayesian inference code has explored the full multidimensional posterior.
    \cite{D_Emilio_2021,Ghosh_2021, Golomb_2022,
        cho2013,GW150914-num,RIFT,
        jaranowski2007gravitationalwave}.
These approximations enable us to better understand gravitational-wave
    sources, and perform hierarchical inference.
These methods include nonparametric approximations such as RIFT
    \cite{RIFT},
    and sample-based estimates such as histograms and 
    Kernel Density Estimates (KDEs)
    \cite{Ghosh_2021}.
In practice, these methods may suffer from performance limitations and their
    own boundary condition limitations.
Histograms in particular, face binning effects and a lack of smoothness.
Better adapted non-parametric methods like carefully tuned
    Gaussian mixture models can significantly mitigate binning
    and smoothness effects;
    see e.g. \cite{Golomb_2022} for an example with GW170817.
The multivariate normal distribution, in particular,
    works well for describing the likelihood function for the
    astrophysical parameters of gravitational-wave events
    \cite{cho2013,GW150914-num,RIFT,jaranowski2007gravitationalwave}.
A truncated set of samples can introduce bias in
    the standard method of implementing the multivariate normal distribution.
Previously, we have introduced optimized fits for
    the bounded multivariate normal distribution in the aligned-spin case,
    which overcome this source of bias
\cite{nal-chieff-paper}.

In this work, we introduce high dimensional bounded (truncated) multivariate
    Normal Approximate Likelihood (NAL) models
    in order to fully characterize precessing spin degrees of freedom
    for gravitational-wave events, as well as
    tidal parameters for binary neutron star mergers.
We provide a detailed algorithm and code to perform the necessary constrained high-dimensional optimization of
    pertinent Gaussian parameters.
We apply our algorithm to the gravitational-wave
    events observed by LIGO-Virgo-Kagra in their first three observing runs,
    finding simple (truncated) Gaussian likelihoods for
    the astrophysical parameters of interest for each event,
    which we report in an associated data release.

Accurate Gaussian likelihood approximations have an enormous range of 
    practical applications in gravitational-wave astrophysics.
For example, because of their simplicity, they're a compact and
    human-comprehensible expression of binary parameters.
They're also an extremely efficient to employ in large-scale
    population inference calculations,
    circumventing many problems associated with using finite sets of 
    posterior samples
    (i.e., the ``delta function'' problem when a
    model avoids any sample;
    and more generally the voracious need for more
    samples from every event as observations are increasingly constraining)
Furthermore, many groups are working to constrain astrophysical models such as
    the evolution of massive stellar binaries and 
    the neutron star equation of state using methods such 
    as Bayesian hierarchical inference
    \cite{1991ApJNarayan, Belczynski2020, broekgaarden2021formation,
    LIGO-O2-RP, LIGO-O3-O3a-RP, LIGO-O3-O3b-RP, Wysocki2018,
    Wysocki2019, Wysocki2020,Rinaldi_2021,D_Emilio_2021,
    Golomb_2022, Breivik_2020,Sadiq2021,Talbot2019GWPopulation,Edelman2021,
    Tiwari_2021,Ghosh_2021,posydon}

Finally, truncated Gaussian approximations 
    have immediate practical utility for low-latency parameter inference
    due to their swift evaluation and compact set of model parameters.

This paper is organized as follows.
In section \ref{sec:methods}, we describe the generation of the
    multivariate normal distribution from the LIGO catalog posterior samples,
    as well as the non-parametric Gaussian Process model.
We justify their validity as approximations to the
    true likelihood describing an observation.
In section \ref{sec:results}, we describe the properties of the likelihood
    models generated for each event.
We summarize the immediate value of these models to other applications.
Finally,
    we summarize the astrophysical consequences of our findings,
    and we infer properties of the underlying compact binary population
    in section \ref{sec:conclude}.

\begin{subsection}{The rationale  behind a simple likelihood approximation}
In brief, our rationale is that current sample-based methods have intrinsic limitations.
Although detailed and accurate approximations exist,
    a Gaussian approximation is fast, easy to interpret,
    and can be made without sacrificing more information than
    is lost through waveform systematics (see Figure \ref{fig:kl}).

Many of the  impactful science goals deduced  from the observed sample of gravitational wave observations can only be
addressed by comparing the full sample of gravitational wave observations $\{ d_k \}$ to a (parametric) population model,  which
predicts the properties $x$ of individual sources as random realizations of a population model $p(x|\Lambda)$ where
$\Lambda$ are population model parameters.   The consistency of any one gravitational wave observation and this
population model can be quantified by a single-event marginal likelihood $\ell_k(\Lambda) = \int dx {\cal
L}_k(x)p(x|\Lambda)$, where the likelihood ${\cal L}_k(x) \propto p(d_k|x)$ quantifies the probability that the observed
data $d_k$ is consistent with Gaussian noise plus a single merger signal with parameters $x$.
However, gravitational wave parameter inference results are currently usually characterized by discrete samples: fair draws from the
posterior distribution  $\propto {\cal L}_x(x)p(x)$ for some fiducial prior \cite{2022arXiv220400461E}.  Quantities like
the marginal likelihood can be evaluated with these samples by by reweighting; see, e.g., Appendix B
in \cite{2018PhRvD..97d3014W}. 

Unfortunately, many GW science objectives involve calculations where continuous inputs, not samples, might be preferable.
These scenarios occur when the population model $p(x|\Lambda)$ has sharp features, such as narrow
subpopulation (e.g., black holes born with exactly zero spin, or neutron stars with a specific equation of state).
When $p(x|\Lambda)$ is very narrow, sample-based estimates for $\ell_k$ break down
\cite{2019RNAAS...3...66F,2022arXiv220400461E,2020arXiv201201317T}.
Worse, as demonstrated by example Appendix B of \cite{gwastro-RIFT-Update}, even ubiquitous spin prior reweighting can
frequently introduce nominally  infinite  variance in pertinent Monte Carlo sums.  

Fortunately, several techniques have been or are being developed to reconstruct the full GW likelihood, or more
typically the likelihood marginalized over several nuisance parameters in $x$ like the sky location and polarization.
RIFT, the most well-developed, uses the marginal likelihood directly in its parameter
inference \cite{gwastro-RIFT-Update,RIFT,gwastro-PE-Code-RIFT}, and the interpolated likelihood output has
been directly used in EOS and population inference \cite{LIGO-GW170817-EOSrank,Wysocki2020,gwastro-nsnuc-SteinerJoint-2020}; see also similar likelihood interpolation
elsewhere for neutron stars
\cite{2020MNRAS.499.5972H}.
Other groups are now
  prototyping methods to accurately reconstruct the likelihood overall \cite{D_Emilio_2021,Golomb_2022}
  and in limited circumstances \cite{2022arXiv221106435R}.

Also, in early low-dimensional investigations using very few observable properties of  $x$, many groups have adopted a
dangerous kernel-density-estimate approach (e.g., \cite{Ghosh_2021,2019PhRvD.100h3015W,2019PhRvD..99h4049L,2021PhRvD.104f2009M,2022arXiv221106435R}), attempting to reconstruct the single-event
(marginal) likelihood over a small number of parameters simply by reweighting and smoothing the available samples.
While viable for a handful of dimensions for typical sets of $\simeq 10^4$ posterior samples, such an approach breaks
down  when working with the
eight (for precessing binary black holes) to ten (for precessing binary neutron stars) dimensions required for generic sources.
Aside from being fairly inaccurate and unable to capture many observations, these KDE-based methods are generally very
slow, with computational cost scaling as the training data size.

A fast, easy-to-interpret approximation to the (marginal) likelihood ${\cal L}$ will be useful exactly insofar as it
enables sufficiently accurate population inferences.  To assess the threshold at which an approximation begins to bias
our population estimates, we consider a one-observable toy model: a Gaussian population of binaries with parameter $x$
(standard deviation $\sigma_{\rm pop}$), probed by Gaussian observations of $x$ (standard deviation $\sigma$).  We will
assume our approximate likelihood model introduces random measurement errors $\delta x$ with zero mean and standard deviation
$\sigma_{\rm sys}$.   In this model, we can measure the sample mean to be consistent with zero (e.g., by analogy to
tests for zero average $\chi_{\rm eff}$) to an accuracy $\sqrt{(\sigma_{pop}^2+\sigma^2+\sigma_{sys}^2)/N}$.  For this
type of measurement, so long as the systematic error is small compared to both the population and individual measurement
error, we'll still draw the  conclusions about the average value of $x$ with the same precision.   Similarly, statements
about the population width are only impacted if the systematic measurement error is comparable to or larger than the
uncertainty in individual measurements or the population width.   In short, unless dramatic, small random errors will
usually average out.

More dangerous is the  possibility that a likelihood approximation technique could be \emph{consistently} biased (i.e.,  the
mean of $\delta x$ is not zero).  For example, a likelihood approximation that was consistently but slightly biased
against equal mass would, over time, produce evidence against all binaries being equal mass, even for a population of
twins.  To contaminate a population measurement with $N$ observations, then, we would require a large bias of $\E{\delta
x} > \sqrt{(\sigma_{pop}^2+\sigma^2)/N}$, equivalent to shifting the mean by of order $1/\sqrt{N}$ standard deviations.
For the near future, with only  $N \lesssim 10^3$ observations available, systematic biases smaller than the equivalent
of a $3\%$ shift in mean will be difficult to identify with poulation-level measurements, and then only if this
hypothetical systematic  produce a consistent impact one parameter for  all events.  [By contrast, waveform systematics often produces
much larger differences, albeit in less-consistent parameter directions; see, e.g.,  
\cite{GWTC-1,GW190412,GWTC-2,GWTC-2p1,GWTC-3}.]  

In this paper, we characterize distribution differences with the KL divergence $D(p|q) = \int p \ln p/q dx
$.  For two Gaussian distributions, a difference of $\sigma/\sqrt{N}$ in the sample mean with minimal shift in variance coresponds to a KL divergence
of $1/2N$.  We therefore adopt $1/2N$ as the most pessimistic scale for observationally-pertinent for KL divergences
between observations, where $N$ is a typical number of observations.

\end{subsection}
 \end{section}

\begin{section}{Methods}
\label{sec:methods}

\begin{figure}[!h]
\centering
\includegraphics[width=3.375 in]{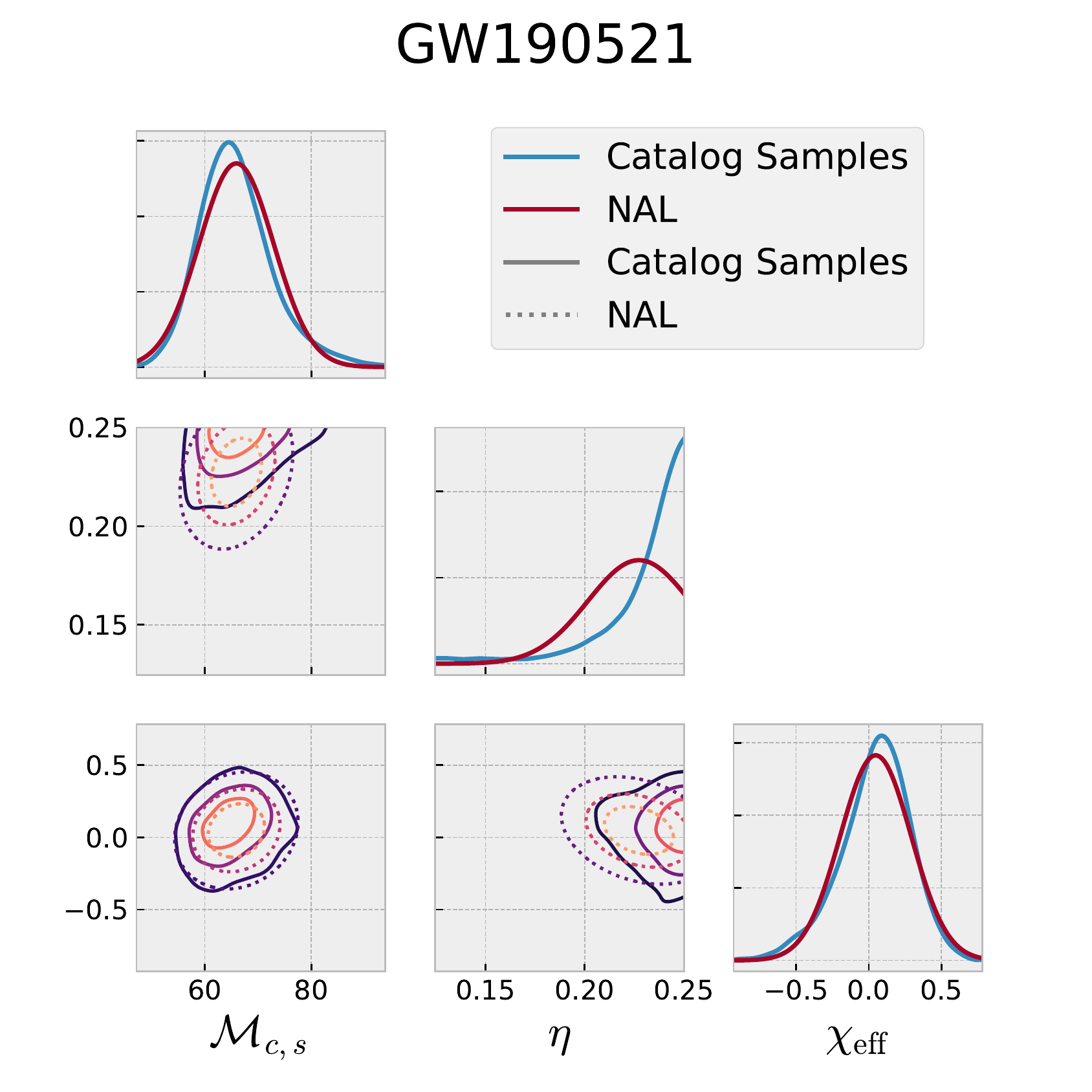}
\newline
\includegraphics[width=3.375 in]{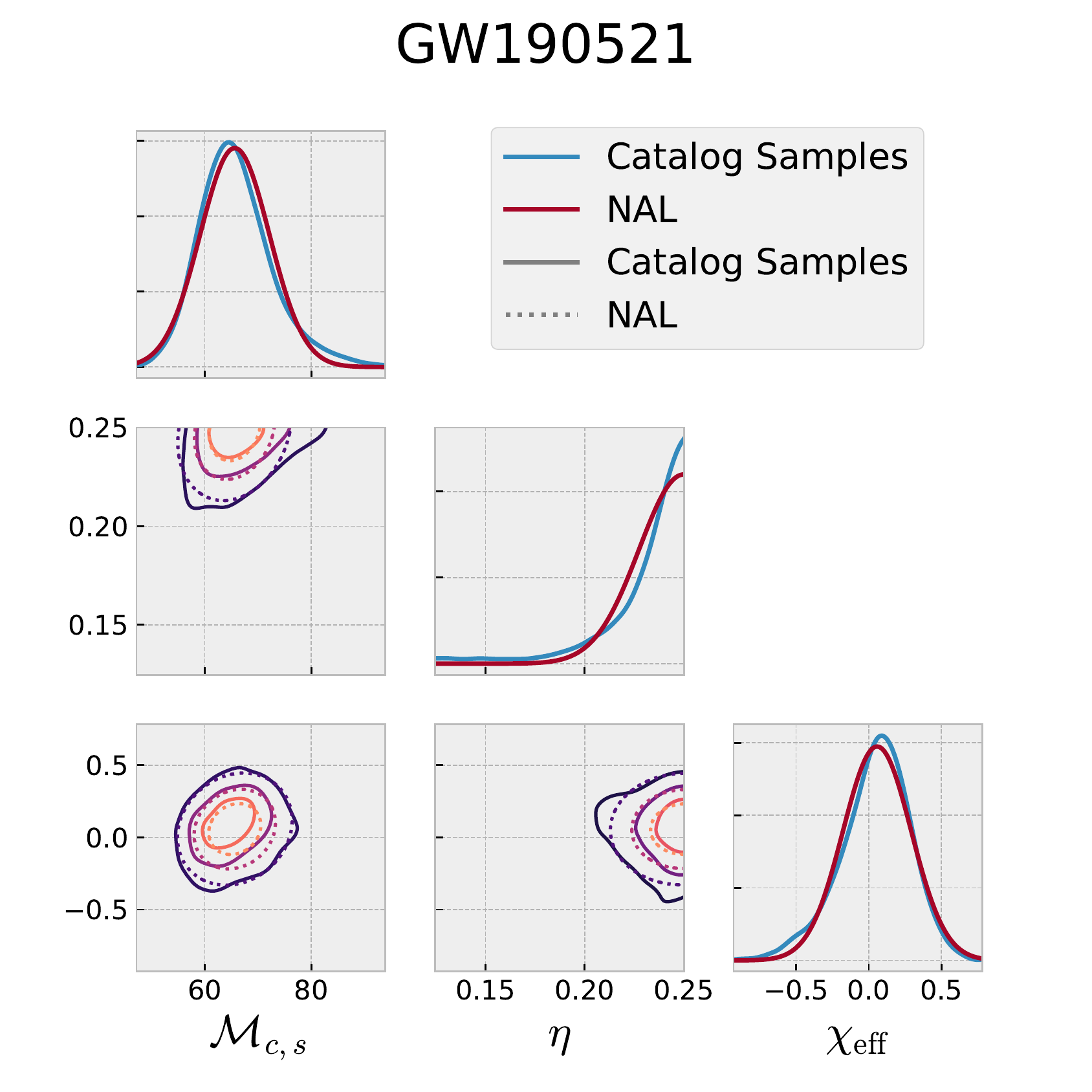}
\caption{\label{fig:justification}
Multivariate normal likelihood approximation using two methods:
(left) implicit parameterization and (right) optimized parameterization.
Diagonal plots are one-dimensional marginalizations of the likelihood
    function, normalized on our bounded interval.
Off-diagonal plots represent two-dimensional marginalizations,
    with contours drawn enclosing $[25, 50, 75]$ confidence intervals.
The PublicationSamples waveform samples for GW190521 are used for this example.
}
\end{figure}

The multivariate normal distribution has long been used for
    gravitational-wave likelihood approximation
    \cite{GW150914-num, PoissonGW1995, RichardPEFisher2014,
        ChoFisher2013, RichardPEPrecessing2014}.
However, many useful coordinates (such as mass ratio $q$ and
    symmetric mass ratio $\eta$), for which gravitational
    wave events are well parameterized, are affected by finite
    boundary effects.
This causes a displacement in the peak of the likelihood function
    from the mean or median of our parameter estimation samples
    (see Figure \ref{fig:justification}).
By optimizing the parameters of the multivariate normal distribution,
    normalized on a finite interval, we can overcome these limitations
    \cite{nal-chieff-paper}.

\begin{subsection}{Bayesian Inference}\label{sec:prior}
Using full numerical relativity to simulate precessing merger events
    is the only way to fully characterize a binary black hole inspiral
    and merger in complete rigor.
While some groups are actively working on this 
    \cite{HealyNR,GW150914-num,Ramos-Buades-NR,Husa-NR,
        NINJA, NRAR, SXS2013, ETK2012, ETK2013,
        Boyle_2019_SxS,Jani_2016_GA_Tech, BAM2008},
    computational limitations impose difficulties on such
    parameter estimation implementations
    \cite{Varma_2019}.
The catalog associated with the first part of LIGO's third observing run (O3a)
    explains that phenomenological, effective one body,
    and numerical relativity surrogate waveform approximants
    are used in place of direct numerical relativity parameter estimation
    for the Gravitational-wave Transient Catalogs
    \cite{GWTC-2}.
We know that this posterior is the
    product $\mathcal{L}(x) \times p(x)$, where $p(x)$ is a fiducial prior, and
    $\mathcal{L}(x)$ is the marginal likelihood that a waveform with parameters $x$
    describes a real gravitational-wave observation.
In this study we include likelihood approximations
    with the parameter estimation samples provided in the
    Gravitational-Wave Transient Catalogs
    \cite{GWTC-1, GWTC-2, GWTC-2p1, GWTC-3, GWTC-2p1-Zenodo, GWTC-3-Zenodo}.
For each event in LIGO--Virgo's third observing run,
    we provide Cartesian-spin models as well as aligned-spin models
    for each of the high modal waveforms available in the publicly
    released samples for each event.
Waveforms included in the associated data release include
    IMRPhenomD \cite{IMRPhenomDa, IMRPhenomDb},
    IMRPhenomPv2 \cite{IMRPhenomPv2, IMRPhenomPv2b},
    IMRPhenomPv3 \cite{IMRPhenomPv3},
    IMRPheomPv3HM \cite{IMRPhenomPv3HM},
    IMRPhenomXPHM \cite{IMRPhenomXPHMa, IMRPhenomXPHMb, IMRPhenomXPHMc},
    SEOBNRv3 \cite{SEOBNRv3a, SEOBNRv3b},
    SEOBNRv4 \cite{SEOBNRv4},
    SEOBNRv4PHM \cite{SEOBNRv4PHMa, SEOBNRv4PHMb},
    and NRSur7dq4 \cite{NRSur7dq4}
    as well as others
    \cite{TEOBResumSa, TEOBResumSb, NRTidalExt, NRTidalv2Ext, SEOBNRv4Ta, SEOBNRv4Tb}.

Developing a likelihood approximation from posterior samples
    without revisting gravitational-wave strain data  
requires re-weighing fair draw posterior samples
    by the inverse prior for each set of astrophysical parameters studied
    \cite{Bayes}.
The mass prior removed from each posterior is uniform in detector component
    masses:
\begin{align}\label{eq:mass-prior}
p(\mathcal{M}_{c,z}, \eta) \mathrm{d}\mathcal{M}_{c,z} \mathrm{d}\eta =
    \frac{4}{(M_{\mathrm{max}} - m_{\mathrm{min}})^2}
    \frac{\mathcal{M}_{c,z}\mathrm{d}\mathcal{M}_{c,z}\mathrm{d}\eta}{
        \eta^{6/5}\sqrt{1 - 4 \eta}}
\end{align}
This is consistent with prior work \cite{RIFT}.
The prior in spin removed for each posterior is uniform in
    Cartesian spin components \cite{CallisterPrior2021}.
For parameterizations which include a distance parameter,
    the prior removed is as the inverse square of the luminosity
    distance.
A uniform prior is assumed for neutron star tidal parameters.

\end{subsection}
\begin{subsection}{Likelihood estimation}

For parameters of which Gaussian noise is expected,
    but that face finite boundary constraints that significantly truncate
    a sampled distribution,
    direct inference of the $\mathbf{\mu}$ and $\mathbf{\Sigma}$
    parameters for a bounded multivariate normal distribution
    (from the sample mean and covariance) is a poor approximation.
As with any parametric model,
    an alternate construction of $\mathbf{\mu}$ and $\mathbf{\Sigma}$
    for a bounded multivariate normal distribution
    is to optimize those model parameters.
This optimization compares the bounded multivariate normal distribution
    $\mathcal{L}(\mathbf{x} | \mathbf{\mu}, \mathbf{\Sigma}) \propto
    G(\mathbf{x} - \mathbf{\mu}, \mathbf{\Sigma})$ to
    an estimate of the 
    marginalized sample distribution for each combination of one and
    two astrophysical parameters of interest
    (normalized on a bounded interval).
To illustrate the effectiveness of this method,
    Figure \ref{fig:justification} shows the benefit of an optimized fit
    visually.
In that example for GW190521,
    the offset in $\mu_{\eta}$ is $1.006 \sigma_{\eta}$
    (from the optimized parameterization),
    where the marginals for the optimized method
    have an average KL divergence of
    $0.016$, compared to the mean and covariance estimate's $0.074$.
This demonstrates a clear increase in the optimized parameterization's
    goodness of fit, over the simple inferred parameterization.
More details about the estimation of this KL divergence 
    statistic are included in Appendix \ref{ap:optimization}.

\begin{subsubsection}{Multivariate normal parameterization}

The Multivariate Normal distribution has many properties which are ideal
    for population models, such as trivial normalization and
    ease of drawing
    fair-draw independent identically distributed random samples.
The peak of the multivariate normal likelihood is characterized by
    a set of parameters $\mathbf{\mu}$.
Its shape is determined by a characteristic covariance
    $\mathbf{\Sigma}$.
The multivariate normal likelihood is given by
\begin{align}\label{eq:norm}
G(\mathbf{x}-\mathbf{\mu}, \mathbf{\Sigma}) =
    (|2\pi \mathbf{\Sigma}|)^{-\frac{1}{2}}
    \exp \Big[
    -\frac{1}{2} (\mathbf{x} - \mathbf{\mu})^T \mathbf{\Sigma}^{-1}(\mathbf{x}-\mathbf{\mu})
    \Big]
\end{align}

Values for each component of $\mathbf{\mu}$ and $\mathbf{\Sigma}$
    can be inferred implicitly from the mean and covariance
    of a sample distribution.
As seen in Figure \ref{fig:justification}, this has limitations
    when samples are affected by a finite boundary condition.
We need to incorporate an assumption that our likelihood function
    $\mathcal{L}(\mathbf{x}) \propto G(\mathbf{x} - \mathbf{\mu},\mathbf{\Sigma})$
    inside of our bounded interval, and that $\mathcal{L}(\mathbf{x}) = 0$
    outside of that interval.

We optimize the multivariate normal distribution as a parametric model
using maximum likelihood estimation (see appendix \ref{ap:optimization}).
The characteristic covariance parameters can be further decomposed
    to reduce degeneracy in our model.
$\mathbf{\Sigma} = \mathbf{\sigma} \mathbf{\rho} \mathbf{\sigma}$,
    where $\mathbf{\sigma}$ is a characteristic standard deviation,
    and $\mathbf{\rho}$ is a characteristic correlation matrix.
The correlation parameters have useful properties of symmetry
    ($\rho_{i,j} = \rho_{j,i}$),
    unity along the diagonal ($\rho_{i,i} = 1$),
    and are bounded within $-1 \le \rho_{i,j} \le 1$.
A multivariate normal distribution with $k$ dimensions
    will have $k$ parameters in $\mathbf{\mu}$,
    $k$ parameters in $\mathbf{\sigma}$,
    and $(k^2 - k)/2$ parameters in $\mathbf{\rho}$.
These total $(k^2 + 3k)/2$.

A fully generic optimization of a parametric model describing the
    sample population would require an evaluation set $\mathbf{X}_{eval}$
    for the full dimensionality of our fit.
As the number of dimensions increases, the number of samples required would
    increase alongside it.
At this stage, we take advantage of the fact that the multivariate normal
    distribution can be trivially marginalized to one- and two-dimensional
    marginals, without any loss of information.
We fit all of the corresponding one- and two-dimensional marginals,
    $G(x_i - \mu_i, \sigma_i)$ and
    $G([x_i,x_j] - [\mu_i,\mu_j],
        [[\sigma_i^2, \sigma_i\sigma_j\rho_{ij}],
         [\sigma_i\sigma_j\rho_{ij}, \sigma_j^2]])$,
    which are independently renormalized in their own bounded regions.
In doing so, we also fit $G(\mathbf{x} - \mathbf{\mu}, \mathbf{\Sigma})$.

The goodness of fit criterion for the NAL models are evaluated using
    a discretized Kullback-Leibler (KL) divergence,
    comparing the NAL approximation of $\mathcal{L}(\mathbf{x})$
    to that of the Gaussian process 
    for each marginalization(see appendix \ref{ap:optimization} for more about error reduction)
    \cite{KL-div}.

\end{subsubsection}

\begin{subsubsection}{Non-parametric likelihood estimation}
\label{sec:nonparametric}

Optimizing the multivariate normal distribution
    as a parametric model
benefits from an intermediate estimation of the marginal sample density.
[The underlying reweighted posterior samples characterize a higher-dimensional posterior than the one we
    seek to model with a Gaussian, with many nuisance parameters; we do not simply fit a multivariate Gaussian
    applied to the whole the
    15-dimensional posterior.]
Instead, we use the fact that any multivariate Gaussian is fully determined by all of its one and
  two-dimensional Gaussian marginal distributions.  With this general feature, we can reconstruct our multivariate
  Gaussian by comparing our Gaussian model to an estimated full marginal likelihood  on these much simpler (and
  numerically very stable) lower-dimensional marginal distributions.

One set of methods for identifying these one- and two-dimensional marginal densities is to use histograms
    with various smoothing and binning procedures that reduce
    the risk of overfitting or underfitting a sample
    \cite{ScargleHistogram2012},
    and many of these solutions are used in gravitational-wave
    likelihood estimation
    \cite{Cornish2010, Zackay2018, Cornish2021, Leslie2021, D_Emilio_2021}.
We note that some of these works
    \cite{RIFT, D_Emilio_2021}
    make use of Gaussian Process Regression to interpolate
    samples of the gravitational-wave likelihood function
    for individual events.

Motivated by these and other prior works, we use Gaussian-process approximations to our (reweighted) low-dimensional
posterior histograms, to produce smooth intermediate non-parametric
    likelihood estimates for our one- and two-dimensional marginal distributions.
Our implementation of Gaussian processes toward this objective
    makes use of one- and two-dimensional histograms 
    with a fixed bin width; 
    see Appendix \ref{ap:optimization} for details.
These marginal fits provide a 
    smooth
    interpolation
    of the likelihood function derived from the catalog samples,
    normalized on our bounded interval.

\end{subsubsection}

\end{subsection}
 \end{section}

\begin{section}{Results}
\label{sec:results}
\begin{figure}
\centering
\includegraphics[width=3.375 in]{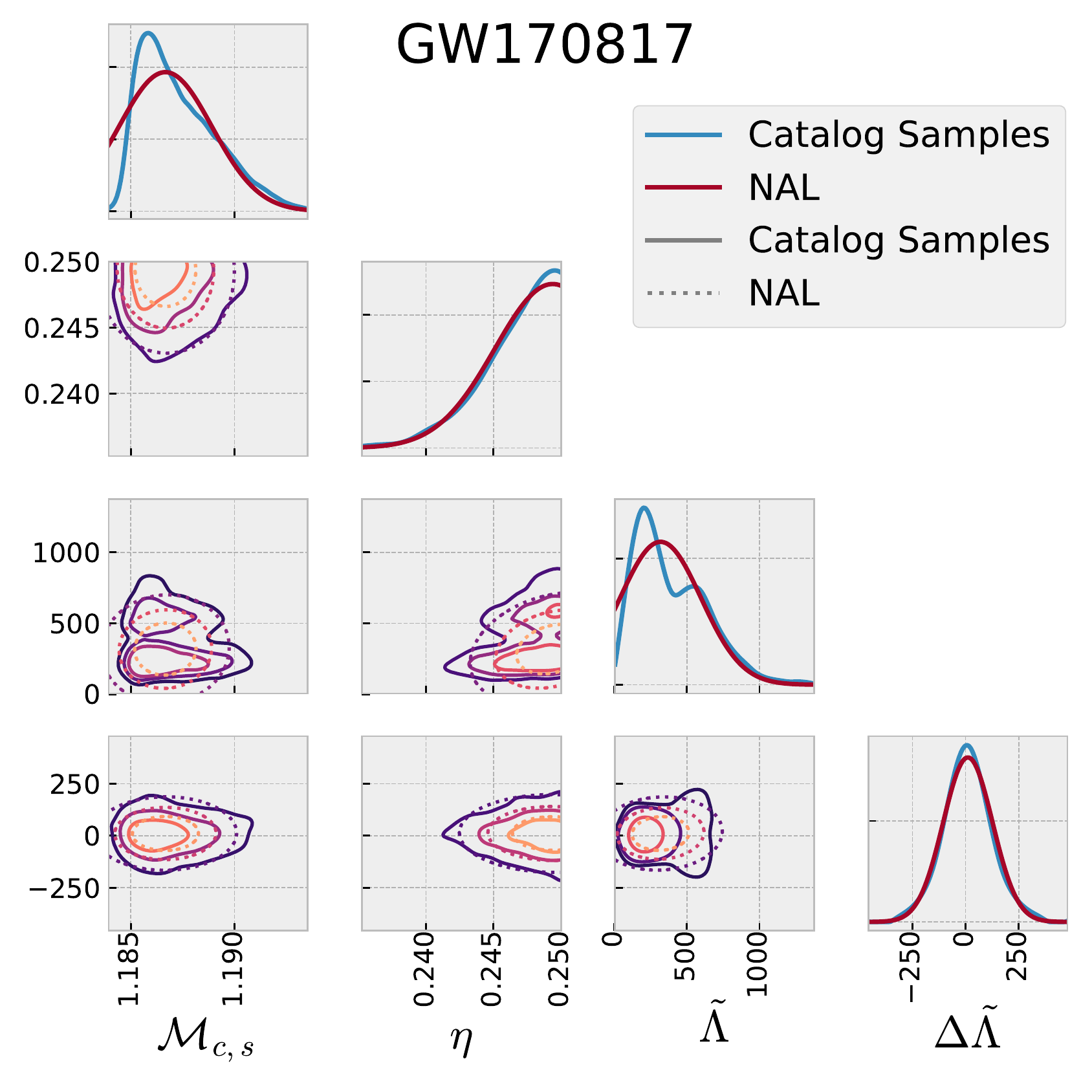}
\caption{\label{fig:tides}
Similar to figure \ref{fig:justification}. Tidal parameters are fit for BNS mergers.
This example includes samples from the IMRPhenomPv2NRT\_lowSpin approximant.
}
\end{figure}

\begin{figure}
\centering
\includegraphics[width=3.375 in]{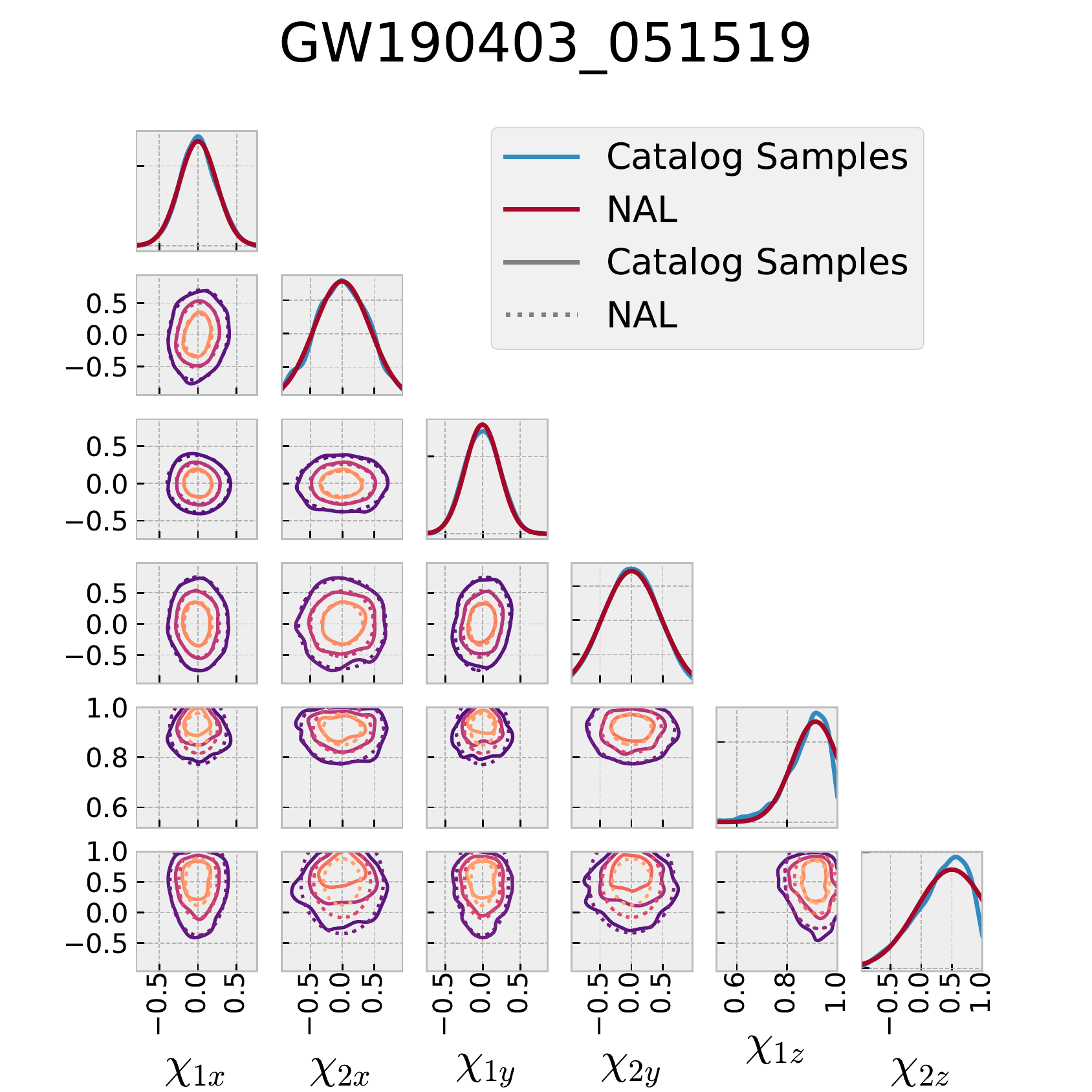}
\caption{\label{fig:precession}
Similar to figure \ref{fig:justification}.
Cartesian spin components are fit, compared with catalog samples.
GW190403\_051519 is interesting because of the high degree of certainty for $\chi_{1z} > 0$,
    known to 10 characteristic standard deviations.
The SEOBRNv4PHM waveform samples for GW190403\_051519 are used for this example.
}
\end{figure}

Figure \ref{fig:kl} highlights the agreement between NAL
    models and the sample distribution for the preferred samples
    of confident BBH events in O3 for both aligned and
    precessing spin parameterizations.
We find that waveform systematic vastly outweigh the information lost
    when NAL models are used to represent the sample distribution.
The KL divergence-based measure of the quality of each  fit is available as part of the
    associated data release;  see Appendix \ref{ap:optimization} for details.
A bounded multivariate normal distribution in these parameters is
    sufficient for fully characterizing the astrophysical parameters of
    events without significant non-Gaussian features (such as bi-modality).
This model has many useful properties,
    including a way to generate random samples,
    fast likelihood evaluation,
    symbolic marginalization and transformations,
    and parameters which describe the Maximum Likelihood Estimate (MLE),
    uncertainty, and parameter correlation.
We provide fits for the mass parameters, distance, aligned spin,
    tidal deformability, and Cartesian spin components for
    available samples in the Gravitational-Wave Transient Catalogs
    \cite{GWTC-1, GWTC-2, GWTC-2p1, GWTC-3, GWTC-2p1-Zenodo, GWTC-3-Zenodo}.

Figure \ref{fig:justification} shows the advantages of these fits
    for three-dimensional likelihood approximations,
    using an aligned spin model.
Figure \ref{fig:tides} shows the agreement for fits to the tidal parameters
    $\tilde{\Lambda}$ and $\delta \tilde{\Lambda}$ with mass parameters.
Tidal parameter fits are available for events/waveform combinations
    with catalog support, including fits for GW170817 and GW190425.
Figure \ref{fig:precession} shows an agreement for Cartesian spin components
    with high spin.
Cartesian spin fits are available for event/waveform combinations
    with catalog support.
Examples for different types of fits are provided in table \ref{tab:summary}

\begin{figure}
\centering
\includegraphics[width=3.375 in]{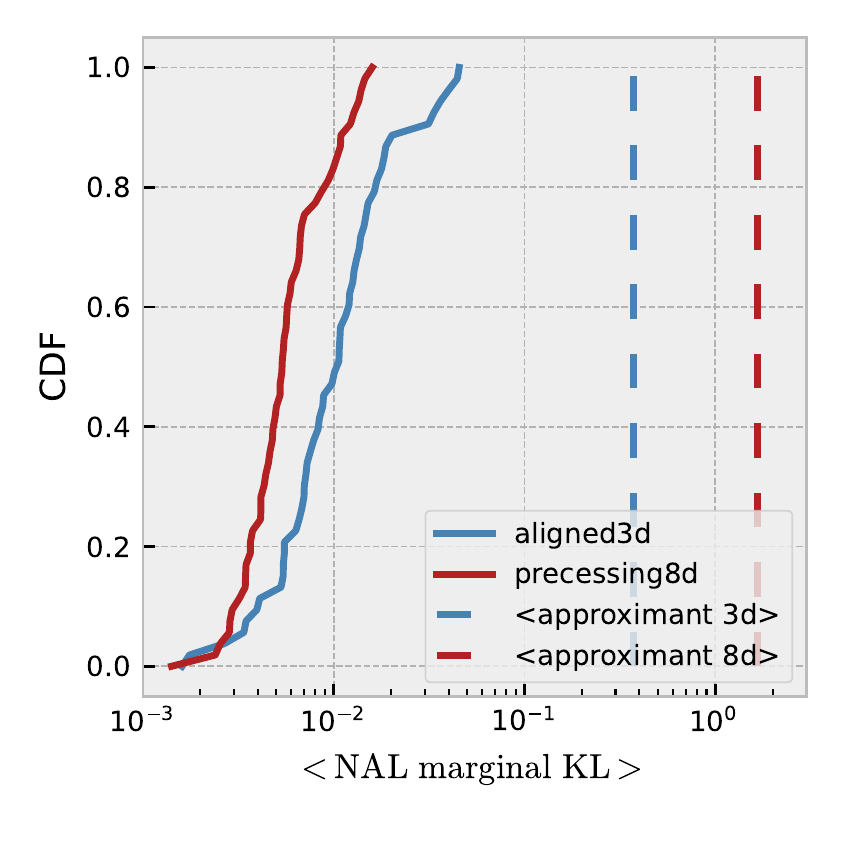}
\caption{\label{fig:kl}
A Cumulative Distribution Function (CDF)
    for the average of the marginal
    KL divergences individual events,
    measuring the agreement between NAL models and
    the sample distribution.
This selection includes the preferred samples for the 53
    confident BBH observations in O3,
    identified by the rates and populations paper
    \cite{LIGO-O3-O3b-RP}.
We consider both aligned and precessing spin
    parameterizations, consistent with table \ref{tab:coord_tags}.
The median KL divergence is comparable to or below the most pessimistic scenario ($1/2N$)
    above which systematics in our likelihood could possibly impact population inference.
Vertical lines further characterize the effects of waveform systematics.
These lines represent
    the average analytic KL divergence (Eq:\ref{eq:analytic_kl})
    between Gaussians constructed from samples of
    different waveform approximants
    (SEOBNRv4PHM and IMRPhenomXPHM for the 22 included
    events where both are available).
}
\end{figure}

\begin{table*}
\centering
\begin{tabular}{|l|c|l|l|}
\hline
Event & Parameters & KL Divergence & KL (simple)\\
\hline \hline
GW190521 &
$\mathcal{M}_c = 65.6^{+6.5}_{-6.5} M_{\odot}$,
$\eta = 0.2498^{+0.0002}_{-0.022}$,
$\chi_{\mathrm{eff}} = 0.057^{+0.23}_{-0.23}$
& 0.016 & 0.071\\
\hline
GW190408\_181802 &
$\mathcal{M}_{c,z} = 23.53^{+0.93}_{-0.94} M_{\odot}$,
$\eta = 0.248^{+0.002}_{-0.017}$,
$\chi_{\mathrm{eff}} = -0.04^{+0.10}_{-0.10}$,
& 0.033 & 0.066\\
&
$\dlum^{-1} = 6.7^{+1.6}_{-1.6} \times 10^{-4} \mathrm{Mpc}^{-1}$
& \\
\hline
GW190425 &
$\mathcal{M}_{c} = 1.438^{+0.013}_{-0.013} M_{\odot}$,
$\eta = 0.227^{+0.018}_{-0.018}$,
$\chi_{\mathrm{eff}} = 0.09^{+0.05}_{-0.05}$,
& 0.019 & 0.037\\
&
$\tilde{\Lambda} = 208^{+872}_{-208}$,
$\delta \tilde{\Lambda} = 37^{+188}_{-37}$
& \\
\hline
GW190517\_054101 &
$\mathcal{M}_{c} = 25.9^{+2.5}_{-2.5} M_{\odot}$,
$\eta = 0.2498^{+0.0002}_{-0.023}$,
& 0.004 & 0.019\\
&
$\chi_{1x} = 0.00^{+0.45}_{-0.45}$,
$\chi_{2x} = 0.00^{+0.48}_{-0.48}$,
$\chi_{1y} = 0.00^{+0.45}_{-0.45}$,
& \\
&
$\chi_{2y} = 0.00^{+0.46}_{-0.46}$,
$\chi_{1z} = 0.66^{+0.19}_{-0.19}$,
$\chi_{2z} = 0.43^{+0.41}_{-0.41}$
& \\
\hline
\end{tabular}
\caption{\label{tab:summary}
Maximum likelihood estimates for choice parameters for a few events,
    using the ``PublicationSamples'' samples for GWTC-2.
Uncertainties represent the characteristic standard deviation of each fit,
    adjusted for coordinate boundaries.
Included are the KL divergences for the optimized parameterizations,
    compared with the simple parameterizations inferred
    from the sample mean and covariance.
}
\end{table*}

\begin{subsection}{Catalog of models}

\begin{figure}
\centering
\includegraphics[width=3.375 in]{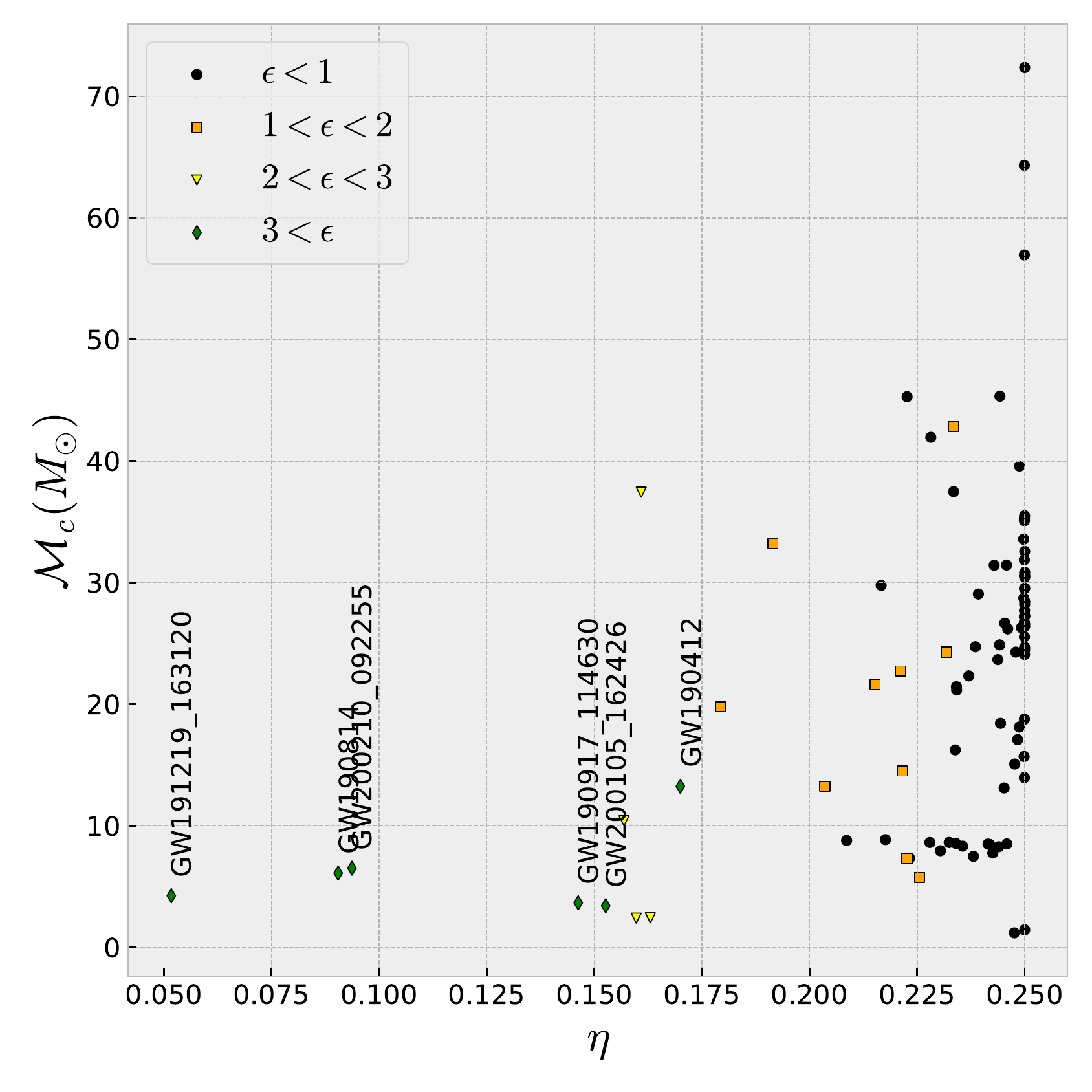}
\caption{\label{fig:scatter_eta_mc}
A scatter plot of MLE parameter values for GW events,
    via aligned3d\_source NAL models, in $\mathcal{M}_c$ and $\eta$.
The markers indicated in the legend designate events
    with statistically significant deviations from equal mass,
    where $\epsilon=(1/4 - \mu_{\eta})/\sigma_{\eta}$.
}
\end{figure}

Figure \ref{fig:scatter_eta_mc} illustrates some properties of
    our NAL fits for the GWTC-3 catalog.
Specifically, this figure provides 
    point estimates for each event based on our inferred 
    Maximum Likelihood Estimate (MLE)
    values for $\mathcal{M}_c$ and $\eta$
    (i.e., it shows our inferred  $\mathbf{\mu}$
    parameters for each event).
Each point estimate is color-coded by a naive frequentist estimate
    of how similar the mass ratio $\eta$ is to equal mass,
    relative to each observation's individual measurement error.  

This figure demonstrates the practical utility of NAL for understanding,
    communicating, and representing GW parameter information:
    we can efficiently characterize individual events'
    properties and highlight events with notably unique properties,
    such as unequal masses.
For example, in this figure,
    the six events color-coded in green are manifestly
    noteworthy as potentially unequal-mass GW sources.
The many familiar frequentist diagnostics enabled by NAL
    should help conceptually identify and differentiate between
    current and future clusters of observations.

As noted previously, our NAL estimates are distinct from and more accurate
    than estimates based on the sample mean or median of 
    parameter estimation samples for events with significant
    effects from a finite boundary in parameter space,
    such as 
is seen in the symmetric mass ratio for 
    events with equal mass.
As demonstrated by Figure  \ref{fig:scatter_eta_mc}, many currently-detectable GW
    sources are  are near or consistent with equal mass.
Other parameters affected by similar finite boundary effects
    include spin ($-1 < \chi_{\mathrm{eff}}$, $\chi_{i,j} <1$  )
    and neutron star deformability ($0 < \tilde{\Lambda}$),
    for which an out-of-bounds parameterization is not physical.
As the shape of the mass distribution, the spin distribution,
    and the neutron star equation of state are key questions
    in this era of gravitational-wave astronomy,
    better  estimates of these parameters are
    of immediate inherent value to studies of the gravitational-wave
    population.

We provide NAL fits for many characterizations of the astrophysical parameters
    for each event in the Gravitational-wave Transient Catalogs,
    for every available waveform approximant in the posterior samples released.
We provide fits for Cartesian spin components and tidal deformability
    parameters for waveform approximants that support those parameters.
Table \ref{tab:coord_tags} (in appendix \ref{ap:params}) 
    outlines the types of fits available
    in our data products, as well as the prior removed from
    each associated set of samples.

\end{subsection}

\section{Applications}
While providing a compact, interpretable representation of GW inferences provides the primary justification
  for using an NAL representation, our approach also opens up additional opportunities.  The following sections briefly
  summarize two opportunities in low-latency parameter inference and in population inference, respectively; see 
\begin{subsection}{Opportunities for low latency parameter estimation}
In the coming observing runs for ground-based gravitational-wave detectors,
    there is a need for low-latency parameterization for the potential follow-up
    of electromagnetic counter-parts to gravitational-wave observations.
Many groups are working to address this challenge,
    using a variety of methods
    \cite{Aasi_2014, Singer_2016, PhysRevLett.127.241103,
        Evans_2012, LIGOLowLatencyUserGuide, Finstad_2020, Morisaki_2020}
Our Gaussian approximations can be efficiently built from sparse training data.
We can train these Gaussians using either posterior samples or marginal likelihoods,
    such as the outputs of the RIFT parameter inference engine.
As a result, NAL can be an ingredient in or an outcome of low-latency parameter inference.
NAL parameterizations also offer an opportunity to interpret posterior sample
    outputs from other low-latency approaches such as the output of a neural net
    (for example, \cite{PhysRevLett.127.241103}).

\begin{figure}
\centering
\includegraphics[width=3.375 in]{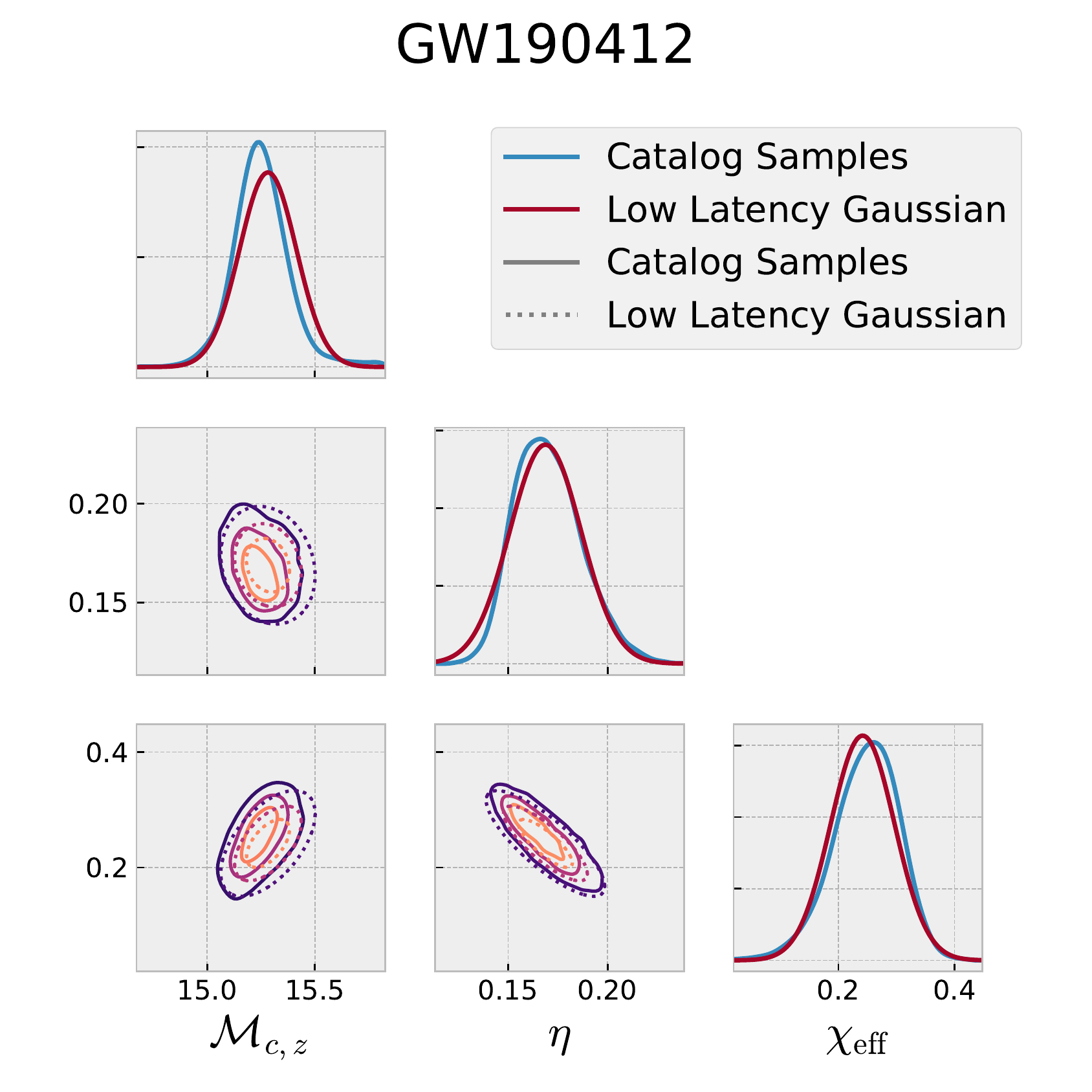}
\caption{\label{fig:lowlatency}
Analysis of GW190412 with NAL based on a limited likelihood inputs 
    from RIFT designed to mimic a
    ``low-latency'' analysis, using IMRPhenomD.
For comparison, 
    we also show the inferred marginal likelihood derived from the discovery
    publication for GW190412, as provided in GWTC-2
    as "PublicationSamples".
Even though the NAL approximation was derived with a different waveform, 
    different physics (no precession), 
    and various different analysis settings than the production results, 
    we find reasonable agreement between this low-latency approach and 
    longer-timescale investigations.
}
\end{figure}
As a concrete example, 
    Figure \ref{fig:lowlatency} shows the results of 
    reconstructing GW190412 with IMRPhenomD,
    based on marginal likelihoods for (detector-frame) 
    masses and spins provided by RIFT on an initial,
    weakly-targeted set of candidate evaluation points in the neighborhood
    of the binary parameters flagged by a search for follow-up.
In this proof-of-concept we employ the detector-frame masses 
    without significant loss of generality,
    because most binary sources meriting follow-up will occur at 
    low redshift and therefore have minimal
    practical difference between their detector and source frame masses.
In this example, a set of points in
    $(\mc,\eta,\chi_{\rm eff})_k$ are evaluated with RIFT to produce a 
    marginal likelihood ${\cal L}$, 
    which we in turn approximate using NAL.
Including queuing and startup time,
    the marginal likelihoods can be evaluated within a few seconds,
    followed by a comparable timescale for 
    the associated Gaussian approximation and posterior.

The NAL result closely conforms to the results of an extended RIFT analysis with comparable settings.
While imperfectly capturing the full range of parameters allowed by more comprehensive physics and waveform systematics,
this fast approximation could be one fruitful method to quickly characterize binary sources, particularly for downstream
rapid population analysis.

We defer a systematic demonstration of and validation study for
    NAL for low-latency PE to a dedicated publication.

\end{subsection}
\begin{subsection}{Population inference}

NAL models can also be efficiently used within population inference calculations,
    to evaluate the necessary event-marginal likelihoods 
    $\int dx \mathcal{L}_n(x)p(x|\Lambda)$ 
    that arise when assessing how well
    a set of individual event likelihoods $\mathcal{L}_n(x)$
    match the predictions of a parameterized population
    $p(x|\Lambda)$,
    where $x$ are binary parameters and $\Lambda$ are population parameters.
The NAL approach provides a fast likelihood and even natural 
    Monte Carlo sampling method for $\mathcal{L}_n$,
    enabling us to efficiently evaluate each of these integrals.

The NAL method is particularly powerful when $p(x|\Lambda)$
    is a generative model only available via Monte Carlo.
Many astrophysical models are formulated by such Monte Carlo techniques,
    notably including isolated binary evolution but
    also many scenarios for dynamical formation
    \cite{Talbot2019GWPopulation, Wysocki2019,
        LIGO-O3-O3b-RP, Belczynski2020,
        Breivik_2020, Sadiq2021, Edelman2021, Tiwari_2021,
        Veitch2015LALInference, Bilby2019}.
In our previous work we have demonstrated how to use these Gaussian
    approximations efficiently with these generative models  
    \cite{nal-chieff-paper,2020ApJ...893...35D}.
For some applications, a direct evaluation of the likelihood is desired,
    such as constraining the neutron star equation of state
    \cite{Ghosh_2021,Golomb_2022,
    Wysocki2020,Rinaldi_2021,D_Emilio_2021}.

The fast likelihood evaluations afforded by NAL compare favorably
    to the speed and accuracy provided by other methods
    particularly in high dimensions.
For instance, a kernel density estimate constructed using the
    fair draws from a posterior or likelihood
    will be slower by a factor that goes as the number of samples
    used to construct the estimate
    (as each sample is represented by a Gaussian 
    for a standard KDE).
This is worsened by the requirement of more and more samples
    for a kernel density estimate as the dimensionality
    of the space explored increases.
When only a fixed number of samples are available,
    there is a limit to the number of dimensions that can be explored
    without losing accuracy.

To illustrate how NAL facilitate efficient comparisons of 
    large population synthesis models to observations,
    Figure \ref{fig:popsyn}, drawn from VD's
    Ph.D. dissertation \cite{DelfaveroDissertation},
    shows the likelihood comparison
    for a one-dimensional binary evolution study,
    focusing on the characteristic value for the Maxwellian
    black-hole kick velocity assumed by the binary evolution model.
The M15 synthetic universe
    (the preferred model with the highest likelihood)
    is based on the M13 simulation from 
    Belczynski et al. (2020)\cite{Belczynski2020}, with the only 
    difference being the
    characteristic value of the Maxwellian black hole kick velocity,
    which takes a value of $130 \mathrm{km/s}$.
This likelihood estimation requires an evaluation of 
    the
    likelihood
    describing the agreement of each simulated binary 
    (for example, M15 consists of 26 million such binaries)
    to each gravitational-wave observation
A kernel density estimate (considering the full set of samples)
    for this event would require an evaluation
    of 26 million Gaussians for each point where you want to evaluate
    the likelihood, while a NAL model requires only one.
This quantifies the speed advantage that NAL models have over
    kernel density estimates in particular,
    and we argue that more complicated methods of evaluating
    an approximate likelihood will not generally be faster
    than a single Gaussian evaluation.
This particular example (figure \ref{fig:popsyn})
    demonstrates this use-case for NAL models.

As the number of events observed by gravitational-wave observatories
    increases, the computational cost of these likelihood evaluations
    becomes increasingly important.
NAL models will readily meet this challenge,
    as a single Gaussian evaluation for each event 
    will cost less than a kernel density estimate for a single event,
    even if the number of events approaches the thousands.
An in-depth discussion of the binary evolution model
    and hierarchical inference involved in this example
    are beyond the scope of this paper,
    but are readily available in the dissertation
    \cite{DelfaveroDissertation}.

\begin{figure}
\centering
\includegraphics[width=3.375 in]{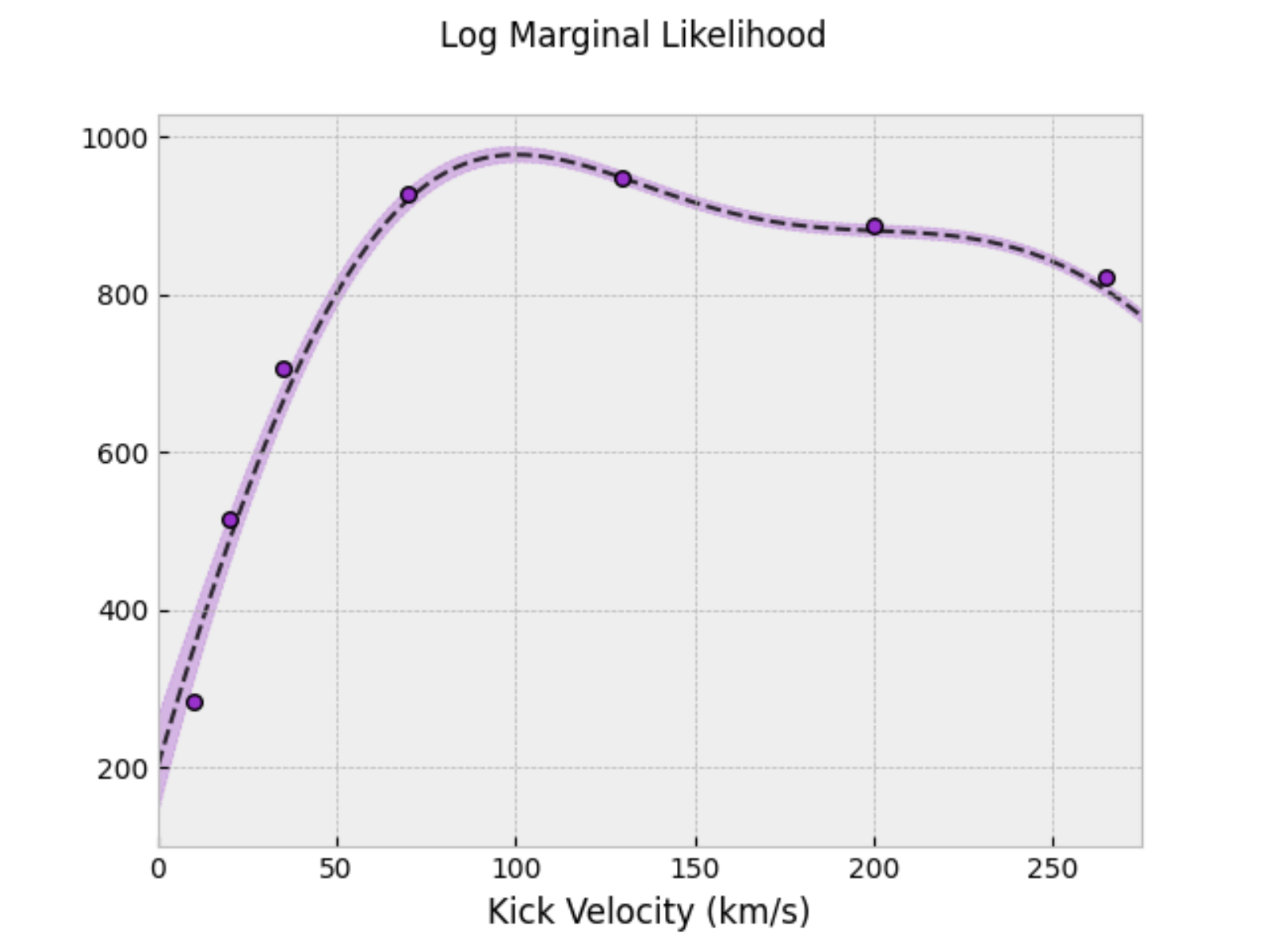}
\caption{\label{fig:popsyn}
Illustration of the utility of NAL:
This population synthesis example
    explores the agreement of binary evolution simulations
    to the observed confident detections in the first part
    of LIGO's third observing run.
The likelihood for each simulation is evaluated using NAL models
    to consider the agreement of each simulated binary to
    each gravitational-wave observation.
This figure originally appeared in previous work
    \cite{DelfaveroDissertation}.
}
\end{figure}

\end{subsection}

 \end{section}

\begin{section}{Conclusions}
\label{sec:conclude}
In this paper, we demonstrate that using NAL models,
    we can take full advantage of the properties of
    the multivariate normal distribution without introducing a bias
    through finite boundary effects.
We demonstrate further that these models succeed in reconstructing
    high dimensional likelihood functions for the parameter estimation
    samples with Cartesian spin components and tidal parameters.
These likelihood models provide the full advantages of the properties of the
    multivariate normal distribution
    (fast evaluation, sampling methods, and symbolic transformations).
Furthermore, the $\mathbf{\mu}$ parameters of each significantly truncated
    Gaussian directly identify the location of maximum likelihood estimate
    more accurately than the sample mean or median.
Therefore, NAL fits provide an advantage for population studies
    wherever the sample mean or median would be employed as the single
    parameter estimate for a gravitational-wave event.

Many other groups estimate the marginal
    likelihood function for each event directly from posterior samples
\cite{D_Emilio_2021,Ghosh_2021, Golomb_2022,
        cho2013,GW150914-num,RIFT,
        jaranowski2007gravitationalwave}.
While these methods have a high degree of accuracy, they suffer from
    computational limitations, especially for higher dimensional
    parameter spaces.
Leading into O4, population studies will be forced to move away from using
    entire catalogs of samples to characterize the likelihood
    for each gravitational-wave event as the sensitivity of the detector
    increases and detections become more abundant
    \cite{fritschel2020instrument}.
This simple parametric approximation to that likelihood function
    provides a reliable tool which will continue to perform well as
    the scale of analysis increases, and already perform well
    to reconstruct the full astrophysical parameter space for
    gravitational-wave events in a computationally efficient way.

At the time of submission of this publication,
    we make the NAL models available publicly at 
    \url{https://gitlab.com/xevra/nal-data},
    as well as the code repository used to make these fits,
    Gravitational-Wave Approximate LiKelihood (GWALK)
    at \url{https://gitlab.com/xevra/gwalk}.
An intermediate step in the construction of NAL models involves using a Gaussian
    Process Regression algorithm to model parameter estimation samples
    \cite{williams2006gaussian}.
Our Gaussian Process code is available at
    \url{https://gitlab.com/xevra/gaussian-process-api}.

 \end{section}

\begin{acknowledgments}
The authors thank Patricia Schmidt for helpful feedback.
ROS, VD, and AY are supported by NSF-PHY 2012057;
    ROS is also supported via NSF PHY-1912632 and AST-1909534.
VD is supported by an appointment to the NASA Postdoctoral Program
    at the NASA Goddard Space Flight Center administered by
    Oak Ridge Associated Universities under contract
    NPP-GSFC-NOV21-0031.
DW thanks the NSF (PHY-1912649) for support.
This material is based upon work supported by NSF’s LIGO Laboratory 
    which is a major facility fully funded by the
    National Science Foundation.
This research has made use of data,
    software and/or web tools obtained from the Gravitational Wave 
    Open Science Center (https://www.gw-openscience.org/ ),
    a service of LIGO Laboratory,
    the LIGO Scientific Collaboration and the Virgo Collaboration.
LIGO Laboratory and Advanced LIGO are funded by the 
    United States National Science Foundation (NSF) as
    well as the Science and Technology Facilities Council (STFC) 
    of the United Kingdom,
    the Max-Planck-Society (MPS), 
    and the State of Niedersachsen/Germany 
    for support of the construction of Advanced LIGO 
    and construction and operation of the GEO600 detector.
Additional support for Advanced LIGO was provided by the 
    Australian Research Council.
Virgo is funded through the European Gravitational Observatory (EGO),
    by the French Centre National de Recherche Scientifique (CNRS),
    the Italian Istituto Nazionale di Fisica Nucleare (INFN),
    and the Dutch Nikhef,
    with contributions by institutions from Belgium, Germany, Greece, Hungary,
    Ireland, Japan, Monaco, Poland, Portugal, Spain.
The authors are grateful for computational resources provided by the 
    LIGO Laboratory and supported by National Science Foundation Grants
    PHY-0757058 and PHY-0823459.
We acknowledge software packages used in producing the 
    gaussian-process-api and GWALK software associated with
    this publication and the associated data release, including
    NUMPY \cite{harris2020array}, SCIPY \cite{2020SciPy-NMeth}, 
    MATPLOTLIB \cite{Hunter_2007}, 
    CYTHON \cite{behnel2011cython},
    ASTROPY \cite{astropy:2013,astropy:2018}, 
    and H5PY \cite{collette_python_hdf5_2014}.
\end{acknowledgments}

\appendix
\begin{section}{Impact of systematics on population inferences}
\label{ap:systematics}
In the introduction, we outlined how systematic errors in our likelihood estimate propagate into population results.  In
this appendix, we provide a more generic treatment.  Our approach applies to \emph{any} systematic, be it deterministic
or stochastic, notably including inaccuracy in the
likelihood estimate (e.g., due to a Gaussian or other approximation; due to limited sample size) and including waveform
systematics.  Since stochastic systematics largely average out, our discussion will emphasize the impact of
deterministic effects.   In the discussion below, our notation will follow Appendix C in \cite{Wysocki2019}.

We consider a systematic controlled by one deterministic parameter $s$.  This parameter could control interpolation between an exact
and a gaussian approximation; waveform systematics; or some other measure of smallness. Each single-event likelihood
realization ${\cal L}(x|s)$ can be expanded in a series in $s$.  For the purposes of this discussion we'll only retain
terms to linear order:
\[
{\cal L}(x|s) \simeq {\cal L}(x|0) + s \partial_s {\cal L} + \ldots
\]
For simplicity omitting selection effects, the population likelihood $L_{\rm pop}$ is the product of single-event
likelihoods
\begin{align}
 L_{\rm pop} = \prod_k {\cal L}_k(x|s) \simeq \left[ \prod_k {\cal L}_k(x|0) \right]
(1+ s \sum_k \frac{\partial_s {\cal     L}_k}{{\cal L}_k} )
\end{align}
To assess the impact of systematics in the limit of many observations, we consider the \emph{population-averaged
  log-likelihood}, relative to a population $p(x|\Lambda)$:
\begin{align}
\E{\ln L_{\rm pop}}_{\rm pop} &\equiv N \int p(x|\Lambda) \ln {\cal L}(x|s) \nonumber \\
&\simeq \ln L_{\rm pop}(0|\Lambda) + s N  \int p(x|\Lambda) \partial_s \ln {\cal L}(x|s)_{s=0}
\end{align}
The first term characterizes the likelihood in the absence of systematics.  In the limit of many observations,  we
assume this expression has a single global maximum, without loss of generality at the origin $\Lambda\simeq 0$.  Expanding the first
term to quadratic order in $\Lambda$ and the second term to linear order as
\begin{align}
\ln L_{\rm pop}(0|\Lambda) \simeq \text{const} - \frac{1}{2} N \hat{\Gamma}_{ab} \Lambda_a \Lambda_b \\
G\equiv \int p(x|\Lambda) \partial_s \ln {\cal L}(x|s)_{s=0}  \simeq \text{const} + \Lambda_a \partial_a G
\end{align}
we can show the population-averaged log likelihood has a global maximum of the form
\begin{align}
\E{\ln L_{\rm pop}}_{\rm pop} &= \text{const} -\frac{1}{2}N \Gamma_{ab} (\Lambda-\Lambda_0)_a (\Lambda-\Lambda_0)_b \\
\Lambda_{0,a} &= s \hat{\Sigma}_{ab} \partial_b G
\end{align}
where $\hat{\Sigma} =\hat{\Gamma}^{-1}$
This final expression for $\Lambda_{0}$  expresses population hyperparameter biases in terms of (derivatives of) $G$,
where $G$ represents population-averaged measurement errors.  In short, we can \emph{calculate} the expected impact of
systematics on population hyperparameter inferences.

While exact results for realistic systematics like waveform uncertainty are computationally inaccessible or stochastic, we
can  capture the essential features of many systematic errors with simple toy models.  Calculations of this form
will reproduce the simple order-of-magnitude estimates described in the introduction.
 \end{section}

\begin{section}{Optimization and Goodness of Fit for Non-parametric and Parametric Models}
\label{ap:optimization}
\begin{subsection}{Non-parametric Likelihood Interpolation}

In this section, we describe how the Gaussian processes which describe
    the one- and two-dimensional marginal distributions are trained,
    and how error is systematically minimized at each step.
Let
    $\mathbf{x}$ refer to a one or two dimensional
    set of parameter space.
We begin by using one- and two-dimensional histograms to model the density
    of the catalog samples.
For some number of bins, $n$, a histogram is constructed,
    representing the posterior samples as a likelihood by weighing each sample
    by the inverse prior.
In this section, we call the centers of each bin $X_n$,
    and the bin values, normalized as a density, $Y_n$.
A Gaussian Process (GP), $f_n(\mathbf{x})$,
    is trained using $X_n$ and $Y_n$,
    with training error estimated from weighted histogram bins
    through the use of binomial fraction estimation:
\begin{align}
\sigma_{\mathrm{hist}} = \sqrt{p \times (1-p)}
\end{align}

These Gaussian processes make
    use of sparse Cholesky decomposition
    and cython compiled basis functions
    \cite{williams2006gaussian}.
We provide this GP as a library at
    \url{https://gitlab.com/xevra/gaussian-process-api}.

We then construct a similar histogram with $n + 1$ bins,
    along with a similar
    GP, $f_{n+1}(\mathbf{x})$, trained using $X_{n+1}$ and $Y_{n+1}$.
We then cross-evaluate $f_n$ and $f_{n+1}$ on each respective training set.

\begin{align}
\delta Y_{n} = \abs{Y_n - f_{n+1}(X_n)} \\
\delta Y_{n + 1} = \abs{Y_{n+1} - f_{n}(X_{n+1})}
\end{align}

We call this the "residual error",
    and We find that by minimizing these errors,
    using the Kolmogorov-Smirnov test \cite{kolmogorov1933sulla},
    we select an optimal number of bins for our histogram.
Once this optimization is complete, we train another GP
    using both sets of training data (for $n$ and $n + 1$).
For this final model, training error is estimated using the
    sum in quadrature of histogram error and the residual error.
This final GP for each marginalization, $f(\mathbf{x})$,
    will therefore have finer resolution than either
    $f_n(\mathbf{x})$ or $f_{n + 1}(\mathbf{x})$.
\end{subsection}
\begin{subsection}{NAL Optimization and Error Handling}

The goodness of fit criterion for the NAL models are evaluated using
    a discretized Kullback-Leibler (KL) divergence,
    comparing the NAL approximation of $\mathcal{L}(\mathbf{x})$
    to that of the Gaussian process 
    for each marginalization, ($f(\mathbf{x})$)
    \cite{KL-div}.
The KL divergence describes the amount of information lost by
    us a distribution Q to describe a
    reference distribution, P.
The KL divergence of a discrete random variable $x$,
    from a space $\mathcal{X}$, is given by:
\begin{align}
\mathrm{KL}(P|Q) = \sum\limits_{x \in \mathcal{X}} p(x) \cdot \mathrm{log} \frac{p(x)}{q(x)}
\end{align}
For two multivariate normal distributions, the KL divergence is known
    analytically:
\begin{align}
\label{eq:analytic_kl}
\mathrm{KL}(P|Q) =
    \frac{1}{2}\Big[
    (\mu_2 - \mu_1)^T \Sigma_2^{-1} (\mu_2 - \mu_1) + \\
    \mathrm{tr}(\Sigma_2^{-1} \Sigma_1) -
    \mathrm{ln}\frac{\abs{\Sigma_1}}{\abs{\Sigma2}} -
    n
    \Big]
\end{align}
A two dimensional multivariate normal distribution offset in $\mu$
    by one $\sigma$ in one dimension will therefore have a KL divergence of
    $0.5$.

The evaluation of 
    the KL divergence
    on each
    NAL model is carried out using
    a grid of 100 points in each dimension inside our bounded region
    (100 points for 1D marginalizations and 
    10,000 points for 2D marginalizations).
We use a maximum likelihood estimate approach in optimizing
    the fit parameters for our higher-dimensional Gaussians,
    taking advantage of the one- and two-dimensional kl divergence information
    available.
The mean of the one- and two dimensional kl divergences is used to
    describe the goodness of fit for each bounded multivariate-normal
    distribution included in the associated data release.

\end{subsection}
 \end{section}

\begin{section}{GWALK Parameterizations}
\label{ap:params}
\begin{table*}
\centering
{
\small
\begin{tabular}{|l|l|l|}
\hline
Parameterization & Coordinates & Prior \\
\hline
aligned3d                   & $\mathcal{M}_{c,z}$, $\eta$, $\chi_{\mathrm{eff}}$ & aligned3d \\
aligned3d\_source           & $\mathcal{M}_{c}$, $\eta$, $\chi_{\mathrm{eff}}$ & aligned3d \\
aligned3d\_dist             & $\mathcal{M}_{c,z}$, $\eta$, $\chi_{\mathrm{eff}}$, $\dlum^{-1}$ & aligned3d\_dist \\
mass\_tides                 & $\mathcal{M}_{c,z}$, $\eta$, $\tilde{\Lambda}$, $\delta \tilde{\Lambda}$ & mass\\
mass\_tides\_source         & $\mathcal{M}_{c}$, $\eta$, $\tilde{\Lambda}$, $\delta \tilde{\Lambda}$ & mass \\
aligned\_tides              & $\mathcal{M}_{c,z}$, $\eta$, $\chi_{\mathrm{eff}}$, $\tilde{\Lambda}$, $\delta \tilde{\Lambda}$ & aligned3d \\
aligned\_tides\_source      & $\mathcal{M}_{c}$, $\eta$, $\chi_{\mathrm{eff}}$, $\tilde{\Lambda}$, $\delta \tilde{\Lambda}$ & aligned3d \\
aligned\_tides\_dist        & $\mathcal{M}_{c,z}$, $\eta$, $\chi_{\mathrm{eff}}$, $\tilde{\Lambda}$, $\delta \tilde{\Lambda}$, $\dlum^{-1}$ & aligned3d\_dist \\
spin6d                      & $\chi_{1x}$, $\chi_{2x}$, $\chi_{1y}$, $\chi_{2y}$, $\chi_{1z}$, $\chi_{2z}$ & precessing8d \\
precessing8d                & $\mathcal{M}_{c,z}$, $\eta$, $\chi_{1x}$, $\chi_{2x}$, $\chi_{1y}$, $\chi_{2y}$, $\chi_{1z}$, $\chi_{2z}$ & precessing8d \\
precessing8d\_source        & $\mathcal{M}_{c}$, $\eta$, $\chi_{1x}$, $\chi_{2x}$, $\chi_{1y}$, $\chi_{2y}$, $\chi_{1z}$, $\chi_{2z}$ & precessing8d \\
precessing8d\_dist          & $\mathcal{M}_{c,z}$, $\eta$, $\chi_{1x}$, $\chi_{2x}$, $\chi_{1y}$, $\chi_{2y}$, $\chi_{1z}$, $\chi_{2z}$, $\dlum^{-1}$ & precessing8d\_dist \\
precessing\_tides\_source   & $\mathcal{M}_{c}$, $\eta$, $\chi_{1x}$, $\chi_{2x}$, $\chi_{1y}$, $\chi_{2y}$, $\chi_{1z}$, $\chi_{2z}$, $\tilde{\Lambda}$, $\delta \tilde{\Lambda}$ & precessing8d \\
full\_precessing\_tides     & $\mathcal{M}_{c,z}$, $\eta$, $\chi_{1x}$, $\chi_{2x}$, $\chi_{1y}$, $\chi_{2y}$, $\chi_{1z}$, $\chi_{2z}$, $\tilde{\Lambda}$, $\delta \tilde{\Lambda}$, $\dlum^{-1}$ & precessing8d\_dist \\
\hline
\end{tabular}
}
\caption{\label{tab:coord_tags}
Here, we define the parameterization labels 
    used in our study, and the associated code/data releases.
}

In table \ref{tab:coord_tags}, we describe the labels
    used in the data release to keep track of the parameters included in each
    type of model.
We further clarify that $\mathcal{M}_{c,z}$ is a detector-frame chirp mass,
    $\mathcal{M}_{c}$ is a source-frame chirp mass,
    $\eta$ is the symmetric mass ratio,
    $\dlum^{-1}$ is the inverse of the luminosity distance,
    $\tilde{\Lambda}$ and $\delta \tilde{\Lambda}$ are the tidal parameters,
    $\chi_{\mathrm{eff}}$ is the aligned spin in the plane of the orbit,
    and
    $\chi_{ij}$ is the spin of the $i$-th object (1 or 2)
        in the $j$ dimension (x, y, or z).
The labeling of the prior which is removed from the posterior samples
    for each fit is corresponds to the simplest parameterization label
    which removes the same prior.
All models must account for the mass prior, and some must account for distance.
The uninformative spin prior is removed for different spin parameterizations
    (see sec \ref{sec:prior}).
\end{table*}
 \end{section}

\bibliographystyle{abbrv}
\footnotesize\bibliography{Bibliography.bib}

\begin{thebibliography}{100}

\bibitem{Aasi_2014}
J.~Aasi, J.~Abadie, B.~P. Abbott, R.~Abbott, and et~al.
\newblock {FIRST} {SEARCHES} {FOR} {OPTICAL} {COUNTERPARTS} {TO}
  {GRAVITATIONAL}-{WAVE} {CANDIDATE} {EVENTS}.
\newblock {\em The Astrophysical Journal Supplement Series}, 211(1):7, feb
  2014.

\bibitem{GW150914-num}
B.~Abbott, R.~Abbott, T.~Abbott, M.~Abernathy, and et~al.
\newblock Directly comparing gw150914 with numerical solutions of einstein’s
  equations for binary black hole coalescence.
\newblock {\em Physical Review D}, 94(6), Sep 2016.

\bibitem{GW150914-detection}
B.~Abbott, R.~Abbott, T.~Abbott, M.~Abernathy, and et~al.
\newblock Gw150914: First results from the search for binary black hole
  coalescence with advanced ligo.
\newblock {\em Physical Review D}, 93(12), Jun 2016.

\bibitem{GWTC-1}
B.~Abbott, R.~Abbott, T.~Abbott, S.~Abraham, and et~al.
\newblock Gwtc-1: a gravitational-wave transient catalog of compact binary
  mergers observed by ligo and virgo during the first and second observing
  runs.
\newblock {\em Physical Review X}, 9(3):031040, 2019.

\bibitem{LIGO-O2-RP}
B.~P. Abbott, R.~Abbott, T.~D. Abbott, S.~Abraham, and et~al.
\newblock Binary black hole population properties inferred from the first and
  second observing runs of advanced ligo and advanced virgo.
\newblock {\em The Astrophysical Journal}, 882(2):L24, Sep 2019.

\bibitem{O3-Detector}
B.~P. Abbott, R.~Abbott, T.~D. Abbott, S.~Abraham, and et~al.
\newblock Prospects for observing and localizing gravitational-wave transients
  with advanced {LIGO}, advanced virgo and {KAGRA}.
\newblock {\em Living Reviews in Relativity}, 23(1), sep 2020.

\bibitem{GWTC-2}
R.~Abbott, T.~Abbott, S.~Abraham, F.~Acernese, and et~al.
\newblock Gwtc-2: Compact binary coalescences observed by ligo and virgo during
  the first half of the third observing run.
\newblock {\em Physical Review X}, 11(2):021053, 2021.

\bibitem{GW150914-astro}
B.~{Abbott et al. (The LIGO Scientific Collaboration and the Virgo
  Collaboration)}.
\newblock {Astrophysical Implications of the Binary Black-hole Merger
  GW150914}.
\newblock {\em \apj}, 818:L22, Feb. 2016.

\bibitem{Virgo}
F.~Acernese, M.~Agathos, K.~Agatsuma, D.~Aisa, and et~al.
\newblock Advanced virgo: a second-generation interferometric gravitational
  wave detector.
\newblock {\em Classical and Quantum Gravity}, 32(2):024001, dec 2014.

\bibitem{TEOBResumSb}
S.~Akcay, S.~Bernuzzi, F.~Messina, A.~Nagar, N.~Ortiz, and P.~Rettegno.
\newblock {Effective-one-body multipolar waveform for tidally interacting
  binary neutron stars up to merger}.
\newblock {\em Phys. Rev. D}, 99(4):044051, 2019.

\bibitem{gwastro-nsnuc-SteinerJoint-2020}
M.~{Al-Mamun}, A.~W. {Steiner}, J.~{N{\"a}ttil{\"a}}, J.~{Lange},
  R.~{O'Shaughnessy}, I.~{Tews}, S.~{Gandolfi}, C.~{Heinke}, and S.~{Han}.
\newblock {Combining Electromagnetic and Gravitational-Wave Constraints on
  Neutron-Star Masses and Radii}.
\newblock {\em \prl}, 126(6):061101, Feb. 2021.

\bibitem{Bilby2019}
G.~Ashton, M.~Hübner, P.~D. Lasky, C.~Talbot, K.~Ackley, S.~Biscoveanu,
  Q.~Chu, A.~Divakarla, P.~J. Easter, B.~Goncharov, F.~H. Vivanco, J.~Harms,
  M.~E. Lower, G.~D. Meadors, D.~Melchor, E.~Payne, M.~D. Pitkin, J.~Powell,
  N.~Sarin, R.~J.~E. Smith, and E.~Thrane.
\newblock Bilby: A user-friendly bayesian inference library for
  gravitational-wave astronomy.
\newblock {\em The Astrophysical Journal Supplement Series}, 241(2):27, apr
  2019.

\bibitem{Kagra}
Y.~Aso, Y.~Michimura, K.~Somiya, M.~Ando, and et~al.
\newblock Interferometer design of the kagra gravitational wave detector.
\newblock {\em Phys. Rev. D}, 88:043007, Aug 2013.

\bibitem{astropy:2018}
{Astropy Collaboration}, A.~M. {Price-Whelan}, B.~M. {Sip{\H{o}}cz}, H.~M.
  {G{\"u}nther}, and et~al.
\newblock {The Astropy Project: Building an Open-science Project and Status of
  the v2.0 Core Package}.
\newblock {\em \aj}, 156(3):123, Sept. 2018.

\bibitem{astropy:2013}
{Astropy Collaboration}, T.~P. {Robitaille}, E.~J. {Tollerud}, P.~{Greenfield},
  and et~al.
\newblock {Astropy: A community Python package for astronomy}.
\newblock {\em \aap}, 558:A33, Oct. 2013.

\bibitem{NINJA}
B.~Aylott, J.~G. Baker, W.~D. Boggs, and M.~B. et~al.
\newblock Testing gravitational-wave searches with numerical relativity
  waveforms: results from the first numerical {INJection} analysis ({NINJA})
  project.
\newblock {\em Classical and Quantum Gravity}, 26(16):165008, aug 2009.

\bibitem{Bayes}
T.~Bayes.
\newblock An essay towards solving a problem in the doctrine of chances.
\newblock {\em Phil. Trans. of the Royal Soc. of London}, 53:370--418, 1763.

\bibitem{behnel2011cython}
S.~Behnel, R.~Bradshaw, C.~Citro, L.~Dalcin, D.~S. Seljebotn, and K.~Smith.
\newblock Cython: The best of both worlds.
\newblock {\em Computing in Science \& Engineering}, 13(2):31--39, 2011.

\bibitem{Belczynski2020}
K.~Belczynski, J.~Klencki, C.~E. Fields, A.~Olejak, and et~al.
\newblock Evolutionary roads leading to low effective spins, high black hole
  masses, and o1/o2 rates for ligo/virgo binary black holes.
\newblock {\em Astronomy \& Astrophysics}, 636:A104, Apr 2020.

\bibitem{2019PASP..131b4503B}
C.~M. {Biwer}, C.~D. {Capano}, S.~{De}, M.~{Cabero}, and et~al.
\newblock {PyCBC Inference: A Python-based Parameter Estimation Toolkit for
  Compact Binary Coalescence Signal}.
\newblock {\em \pasp}, 131(996):024503, Feb. 2019.

\bibitem{SEOBNRv4}
A.~Boh\'e et~al.
\newblock {Improved effective-one-body model of spinning, nonprecessing binary
  black holes for the era of gravitational-wave astrophysics with advanced
  detectors}.
\newblock {\em Phys. Rev. D}, 95(4):044028, 2017.

\bibitem{Boyle_2019_SxS}
M.~Boyle, D.~Hemberger, D.~A.~B. Iozzo, G.~Lovelace, and et~al.
\newblock The {SXS} collaboration catalog of binary black hole simulations.
\newblock {\em Classical and Quantum Gravity}, 36(19):195006, sep 2019.

\bibitem{Breivik_2020}
K.~Breivik, S.~Coughlin, M.~Zevin, C.~L. Rodriguez, K.~Kremer, C.~S. Ye, J.~J.
  Andrews, M.~Kurkowski, M.~C. Digman, S.~L. Larson, and F.~A. Rasio.
\newblock {COSMIC} variance in binary population synthesis.
\newblock {\em The Astrophysical Journal}, 898(1):71, jul 2020.

\bibitem{broekgaarden2021formation}
F.~S. Broekgaarden and E.~Berger.
\newblock Formation of the first two black hole--neutron star mergers (gw200115
  and gw200105) from isolated binary evolution.
\newblock {\em The Astrophysical Journal Letters}, 920(1):L13, 2021.

\bibitem{BAM2008}
B.~Brügmann, J.~A. Gonz{\'{a}}lez, M.~Hannam, S.~Husa, U.~Sperhake, and
  W.~Tichy.
\newblock Calibration of moving puncture simulations.
\newblock {\em Physical Review D}, 77(2), jan 2008.

\bibitem{CallisterPrior2021}
T.~A. {Callister}.
\newblock {A Thesaurus for Common Priors in Gravitational-Wave Astronomy}.
\newblock {\em arXiv e-prints}, page arXiv:2104.09508, Apr. 2021.

\bibitem{ChoFisher2013}
H.~{Cho}, E.~{Ochsner}, R.~{O'Shaughnessy}, C.~{Kim}, and C.~{Lee}.
\newblock {Gravitational waves from BH-NS binaries: Phenomenological Fisher
  matricies and parameter estimation using higher harmonics}.
\newblock {\em \prd}, 87:02400--+, Jan. 2013.

\bibitem{cho2013}
H.-S. Cho, E.~Ochsner, R.~O’Shaughnessy, C.~Kim, and C.-H. Lee.
\newblock Gravitational waves from black hole-neutron star binaries: Effective
  fisher matrices and parameter estimation using higher harmonics.
\newblock {\em Physical Review D}, 87(2), Jan 2013.

\bibitem{LIGOLowLatencyUserGuide}
L.~S. Collaboration and V.~Collaboration.
\newblock Ligo/virgo public alerts user guide.
\newblock 2019.

\bibitem{GWTC-2p1-Zenodo}
L.~S. Collaboration and V.~Collaboration.
\newblock {GWTC-2.1: Deep Extended Catalog of Compact Binary Coalescences
  Observed by LIGO and Virgo During the First Half of the Third Observing Run -
  Parameter Estimation Data Release}, July 2021.

\bibitem{GWTC-3-Zenodo}
L.~S. Collaboration, V.~Collaboration, and K.~Collaboration.
\newblock {GWTC-3: Compact Binary Coalescences Observed by LIGO and Virgo
  During the Second Part of the Third Observing Run — Parameter estimation
  data release}, Nov. 2021.

\bibitem{GWTC-3}
T.~L.~S. Collaboration, the Virgo~Collaboration, and the KAGRA Collaboration~et
  al.
\newblock Gwtc-3: Compact binary coalescences observed by ligo and virgo during
  the second part of the third observing run, 2021.

\bibitem{LIGO-O3-O3b-RP}
T.~L.~S. Collaboration, the Virgo~Collaboration, and the KAGRA Collaboration~et
  al.
\newblock The population of merging compact binaries inferred using
  gravitational waves through gwtc-3, 2021.

\bibitem{GWTC-2p1}
T.~L.~S. Collaboration and the Virgo Collaboration~et al.
\newblock Gwtc-2.1: Deep extended catalog of compact binary coalescences
  observed by ligo and virgo during the first half of the third observing run,
  2021.

\bibitem{collette_python_hdf5_2014}
A.~Collette.
\newblock {\em Python and HDF5}.
\newblock O'Reilly, 2013.

\bibitem{Cornish2010}
N.~J. Cornish.
\newblock Fast fisher matrices and lazy likelihoods, 2010.

\bibitem{Cornish2021}
N.~J. Cornish.
\newblock Heterodyned likelihood for rapid gravitational wave parameter
  inference.
\newblock {\em Physical Review D}, 104(10), nov 2021.

\bibitem{SEOBNRv4PHMa}
R.~Cotesta, A.~Buonanno, A.~Boh\'e, A.~Taracchini, I.~Hinder, and S.~Ossokine.
\newblock {Enriching the Symphony of Gravitational Waves from Binary Black
  Holes by Tuning Higher Harmonics}.
\newblock {\em Phys. Rev. D}, 98(8):084028, 2018.

\bibitem{PhysRevLett.127.241103}
M.~Dax, S.~R. Green, J.~Gair, J.~H. Macke, A.~Buonanno, and B.~Sch\"olkopf.
\newblock Real-time gravitational wave science with neural posterior
  estimation.
\newblock {\em Phys. Rev. Lett.}, 127:241103, Dec 2021.

\bibitem{DelfaveroDissertation}
V.~Del~Favero.
\newblock Constraints on compact binary formation and effective gravitational
  wave likelihood approximation.
\newblock 2022.

\bibitem{nal-chieff-paper}
V.~Delfavero, R.~O'Shaughnessy, D.~Wysocki, and A.~Yelikar.
\newblock Normal approximate likelihoods to gravitational wave events.
\newblock 2021.

\bibitem{D_Emilio_2021}
V.~D'Emilio, R.~Green, and V.~Raymond.
\newblock Density estimation with gaussian processes for gravitational wave
  posteriors.
\newblock {\em Monthly Notices of the Royal Astronomical Society},
  508(2):2090--2097, sep 2021.

\bibitem{NRTidalExt}
T.~Dietrich, S.~Bernuzzi, and W.~Tichy.
\newblock {Closed-form tidal approximants for binary neutron star gravitational
  waveforms constructed from high-resolution numerical relativity simulations}.
\newblock {\em Phys. Rev. D}, 96(12):121501, 2017.

\bibitem{NRTidalv2Ext}
T.~Dietrich, A.~Samajdar, S.~Khan, N.~K. Johnson-McDaniel, R.~Dudi, and
  W.~Tichy.
\newblock {Improving the NRTidal model for binary neutron star systems}.
\newblock {\em Phys. Rev. D}, 100(4):044003, 2019.

\bibitem{2020ApJ...893...35D}
Z.~{Doctor}, D.~{Wysocki}, R.~{O'Shaughnessy}, D.~E. {Holz}, and B.~{Farr}.
\newblock {Black Hole Coagulation: Modeling Hierarchical Mergers in Black Hole
  Populations}.
\newblock {\em \apj}, 893(1):35, Apr. 2020.

\bibitem{Edelman2021}
B.~Edelman, F.~J. Rivera-Paleo, J.~D. Merritt, B.~Farr, and et~al.
\newblock Constraining unmodeled physics with compact binary mergers from
  gwtc-1.
\newblock {\em Phys. Rev. D}, 103:042004, Feb 2021.

\bibitem{2022arXiv220400461E}
R.~{Essick} and W.~{Farr}.
\newblock {Precision Requirements for Monte Carlo Sums within Hierarchical
  Bayesian Inference}.
\newblock {\em arXiv e-prints}, page arXiv:2204.00461, Apr. 2022.

\bibitem{Evans_2012}
P.~A. Evans, J.~K. Fridriksson, N.~Gehrels, J.~Homan, and et~al.
\newblock {SWIFT} {FOLLOW}-{UP} {OBSERVATIONS} {OF} {CANDIDATE}
  {GRAVITATIONAL}-{WAVE} {TRANSIENT} {EVENTS}.
\newblock {\em The Astrophysical Journal Supplement Series}, 203(2):28, nov
  2012.

\bibitem{2019RNAAS...3...66F}
W.~M. {Farr}.
\newblock {Accuracy Requirements for Empirically Measured Selection Functions}.
\newblock {\em Research Notes of the American Astronomical Society}, 3(5):66,
  May 2019.

\bibitem{Finstad_2020}
D.~Finstad and D.~A. Brown.
\newblock Fast parameter estimation of binary mergers for multimessenger
  follow-up.
\newblock {\em The Astrophysical Journal}, 905(1):L9, dec 2020.

\bibitem{posydon}
T.~Fragos, J.~J. Andrews, S.~S. Bavera, C.~P.~L. Berry, S.~Coughlin, A.~Dotter,
  P.~Giri, V.~Kalogera, A.~Katsaggelos, K.~Kovlakas, S.~Lalvani, D.~Misra,
  P.~M. Srivastava, Y.~Qin, K.~A. Rocha, J.~Roman-Garza, J.~G. Serra,
  P.~Stahle, M.~Sun, X.~Teng, G.~Trajcevski, N.~H. Tran, Z.~Xing, E.~Zapartas,
  and M.~Zevin.
\newblock Posydon: A general-purpose population synthesis code with detailed
  binary-evolution simulations, 2022.

\bibitem{fritschel2020instrument}
P.~Fritschel, L.~S. Collaboration, et~al.
\newblock Instrument science white paper 2020.
\newblock {\em Technical Report LIGO-T2000407-v3}, 2020.

\bibitem{IMRPhenomXPHMb}
C.~Garc\'\i{}a-Quir\'os, M.~Colleoni, S.~Husa, H.~Estell\'es, G.~Pratten,
  A.~Ramos-Buades, M.~Mateu-Lucena, and R.~Jaume.
\newblock {Multimode frequency-domain model for the gravitational wave signal
  from nonprecessing black-hole binaries}.
\newblock {\em Phys. Rev. D}, 102(6):064002, 2020.

\bibitem{Ghosh_2021}
S.~Ghosh, X.~Liu, J.~Creighton, I.~M.~n. Hernandez, W.~Kastaun, and G.~Pratten.
\newblock Rapid model comparison of equations of state from gravitational wave
  observation of binary neutron star coalescences.
\newblock {\em Phys. Rev. D}, 104:083003, Oct 2021.

\bibitem{Golomb_2022}
J.~Golomb and C.~Talbot.
\newblock Hierarchical inference of binary neutron star mass distribution and
  equation of state with gravitational waves.
\newblock {\em The Astrophysical Journal}, 926(1):79, feb 2022.

\bibitem{IMRPhenomPv2}
M.~Hannam, P.~Schmidt, A.~Boh\'e, L.~Haegel, S.~Husa, F.~Ohme, G.~Pratten, and
  M.~P\"urrer.
\newblock {Simple Model of Complete Precessing Black-Hole-Binary Gravitational
  Waveforms}.
\newblock {\em Phys. Rev. Lett.}, 113(15):151101, 2014.

\bibitem{harris2020array}
C.~R. Harris, K.~J. Millman, S.~J. van~der Walt, R.~Gommers, and et~al.
\newblock Array programming with {NumPy}.
\newblock {\em Nature}, 585(7825):357--362, Sept. 2020.

\bibitem{HealyNR}
J.~{Healy}, C.~O. {Lousto}, J.~{Lange}, and R.~{O'Shaughnessy}.
\newblock {Application of the third RIT binary black hole simulations catalog
  to parameter estimation of gravitational-wave signals from the LIGO-Virgo O1
  and O2 observational runs}.
\newblock {\em \prd}, 102(12):124053, Dec. 2020.

\bibitem{2020MNRAS.499.5972H}
F.~{Hernandez Vivanco}, R.~{Smith}, E.~{Thrane}, and P.~D. {Lasky}.
\newblock {A scalable random forest regressor for combining neutron-star
  equation of state measurements: a case study with GW170817 and GW190425}.
\newblock {\em \mnras}, 499(4):5972--5977, Dec. 2020.

\bibitem{NRAR}
I.~Hinder, A.~Buonanno, M.~Boyle, and Z.~B.~E. et~al.
\newblock Error-analysis and comparison to analytical models of numerical
  waveforms produced by the {NRAR} collaboration.
\newblock {\em Classical and Quantum Gravity}, 31(2):025012, jan 2013.

\bibitem{SEOBNRv4Ta}
T.~Hinderer et~al.
\newblock {Effects of neutron-star dynamic tides on gravitational waveforms
  within the effective-one-body approach}.
\newblock {\em Phys. Rev. Lett.}, 116(18):181101, 2016.

\bibitem{IAS}
Y.~{Huang}, C.-J. {Haster}, J.~{Roulet}, S.~{Vitale}, and et~al.
\newblock {Source properties of the lowest signal-to-noise-ratio binary black
  hole detections}.
\newblock {\em \prd}, 102(10):103024, Nov. 2020.

\bibitem{Hunter_2007}
J.~D. Hunter.
\newblock Matplotlib: A 2d graphics environment.
\newblock {\em Computing in Science \& Engineering}, 9(3):90--95, 2007.

\bibitem{Husa-NR}
S.~Husa, S.~Khan, M.~Hannam, M.~P\"urrer, F.~Ohme, X.~J. Forteza, and
  A.~Boh\'e.
\newblock Frequency-domain gravitational waves from nonprecessing black-hole
  binaries. i. new numerical waveforms and anatomy of the signal.
\newblock {\em Phys. Rev. D}, 93:044006, Feb 2016.

\bibitem{IMRPhenomDa}
S.~Husa, S.~Khan, M.~Hannam, M.~P\"urrer, F.~Ohme, X.~Jim\'enez~Forteza, and
  A.~Boh\'e.
\newblock {Frequency-domain gravitational waves from nonprecessing black-hole
  binaries. I. New numerical waveforms and anatomy of the signal}.
\newblock {\em Phys. Rev. D}, 93(4):044006, 2016.

\bibitem{Jani_2016_GA_Tech}
K.~Jani, J.~Healy, J.~A. Clark, L.~London, P.~Laguna, and D.~Shoemaker.
\newblock Georgia tech catalog of gravitational waveforms.
\newblock {\em Classical and Quantum Gravity}, 33(20):204001, sep 2016.

\bibitem{jaranowski2007gravitationalwave}
P.~Jaranowski and A.~Królak.
\newblock Gravitational-wave data analysis. formalism and sample applications:
  The gaussian case, 2007.

\bibitem{IMRPhenomPv2b}
S.~Khan, K.~Chatziioannou, M.~Hannam, and F.~Ohme.
\newblock Phenomenological model for the gravitational-wave signal from
  precessing binary black holes with two-spin effects.
\newblock {\em Physical Review D}, 100(2), jul 2019.

\bibitem{IMRPhenomDb}
S.~Khan, S.~Husa, M.~Hannam, F.~Ohme, M.~P\"urrer, X.~J. Forteza, and
  A.~Boh\'e.
\newblock Frequency-domain gravitational waves from nonprecessing black-hole
  binaries. ii. a phenomenological model for the advanced detector era.
\newblock {\em Phys. Rev. D}, 93:044007, Feb 2016.

\bibitem{IMRPhenomPv3}
S.~Khan, F.~Ohme, K.~Chatziioannou, and M.~Hannam.
\newblock {Including higher order multipoles in gravitational-wave models for
  precessing binary black holes}.
\newblock {\em Phys. Rev. D}, 101(2):024056, 2020.

\bibitem{kolmogorov1933sulla}
A.~Kolmogorov.
\newblock Sulla determinazione empirica di una lgge di distribuzione.
\newblock {\em Inst. Ital. Attuari, Giorn.}, 4:83--91, 1933.

\bibitem{KL-div}
Kullback.
\newblock {\em Information theory and statistics}.
\newblock John Wiley and Sons, NY, 1959.

\bibitem{2019PhRvD..99h4049L}
P.~{Landry} and R.~{Essick}.
\newblock {Nonparametric inference of the neutron star equation of state from
  gravitational wave observations}.
\newblock {\em \prd}, 99(8):084049, Apr. 2019.

\bibitem{RIFT}
J.~Lange, R.~O'Shaughnessy, and M.~Rizzo.
\newblock Rapid and accurate parameter inference for coalescing, precessing
  compact binaries, 2018.

\bibitem{Leslie2021}
N.~Leslie, L.~Dai, and G.~Pratten.
\newblock Mode-by-mode relative binning: Fast likelihood estimation for
  gravitational waveforms with spin-orbit precession and multiple harmonics.
\newblock {\em Physical Review D}, 104(12), dec 2021.

\bibitem{ETK2012}
F.~Löffler, J.~Faber, E.~Bentivegna, T.~Bode, P.~Diener, R.~Haas, I.~Hinder,
  B.~C. Mundim, C.~D. Ott, E.~Schnetter, G.~Allen, M.~Campanelli, and
  P.~Laguna.
\newblock The einstein toolkit: a community computational infrastructure for
  relativistic astrophysics.
\newblock {\em Classical and Quantum Gravity}, 29(11):115001, may 2012.

\bibitem{2021PhRvD.104f2009M}
S.~{Mastrogiovanni}, K.~{Leyde}, C.~{Karathanasis}, E.~{Chassande-Mottin},
  D.~A. {Steer}, J.~{Gair}, A.~{Ghosh}, R.~{Gray}, S.~{Mukherjee}, and
  S.~{Rinaldi}.
\newblock {On the importance of source population models for gravitational-wave
  cosmology}.
\newblock {\em \prd}, 104(6):062009, Sept. 2021.

\bibitem{Morisaki_2020}
S.~Morisaki and V.~Raymond.
\newblock Rapid parameter estimation of gravitational waves from binary neutron
  star coalescence using focused reduced order quadrature.
\newblock {\em Physical Review D}, 102(10), nov 2020.

\bibitem{SXS2013}
A.~H. Mrou{\'{e} }, M.~A. Scheel, B.~Szil{\'{a}}gyi, H.~P. Pfeiffer, M.~Boyle,
  D.~A. Hemberger, L.~E. Kidder, G.~Lovelace, S.~Ossokine, N.~W. Taylor,
  A.~Zengino{\u{g}}lu, L.~T. Buchman, T.~Chu, E.~Foley, M.~Giesler, R.~Owen,
  and S.~A. Teukolsky.
\newblock Catalog of 174 binary black hole simulations for gravitational wave
  astronomy.
\newblock {\em Physical Review Letters}, 111(24), dec 2013.

\bibitem{TEOBResumSa}
A.~Nagar et~al.
\newblock {Time-domain effective-one-body gravitational waveforms for
  coalescing compact binaries with nonprecessing spins, tides and self-spin
  effects}.
\newblock {\em Phys. Rev. D}, 98(10):104052, 2018.

\bibitem{1991ApJNarayan}
R.~{Narayan}, T.~{Piran}, and A.~{Shemi}.
\newblock {Neutron Star and Black Hole Binaries in the Galaxy}.
\newblock 379:L17, Sept. 1991.

\bibitem{3-OGC}
A.~H. Nitz, C.~D. Capano, S.~Kumar, Y.-F. Wang, S.~Kastha, M.~Schäfer,
  R.~Dhurkunde, and M.~Cabero.
\newblock 3-{OGC}: Catalog of gravitational waves from compact-binary mergers.
\newblock {\em The Astrophysical Journal}, 922(1):76, nov 2021.

\bibitem{RichardPEFisher2014}
R.~{O'Shaughnessy}, B.~{Farr}, E.~{Ochsner}, H.-S. {Cho}, C.~{Kim}, and C.-H.
  {Lee}.
\newblock {Parameter estimation of gravitational waves from nonprecessing black
  hole-neutron star inspirals with higher harmonics: Comparing Markov-chain
  Monte Carlo posteriors to an effective Fisher matrix}.
\newblock {\em \prd}, 89(6):064048, Mar. 2014.

\bibitem{RichardPEPrecessing2014}
R.~{O'Shaughnessy}, B.~{Farr}, E.~{Ochsner}, H.-S. {Cho}, V.~{Raymond},
  C.~{Kim}, and C.-H. {Lee}.
\newblock Parameter estimation of gravitational waves from precessing black
  hole-neutron star inspirals with higher harmonics.
\newblock {\em \prd}, 89:102005, May 2014.

\bibitem{gwastro-PE-Code-RIFT}
R.~{O'Shaughnessy} and J.~{Lange}.
\newblock {Code repository for RIFT: Rapid Iterative FiTting for gravitational
  wave source parameter inference}.
\newblock {\em Available at https://git.ligo.org/rapidpe-rift/rift/}, 2015.

\bibitem{IMRPhenomPv3HM}
S.~Ossokine, A.~Buonanno, S.~Marsat, R.~Cotesta, S.~Babak, T.~Dietrich,
  R.~Haas, I.~Hinder, H.~P. Pfeiffer, M.~Pürrer, C.~J. Woodford, M.~Boyle,
  L.~E. Kidder, M.~A. Scheel, and B.~Szil{\'{a}}gyi.
\newblock Multipolar effective-one-body waveforms for precessing binary black
  holes: Construction and validation.
\newblock {\em Physical Review D}, 102(4), aug 2020.

\bibitem{SEOBNRv4PHMb}
S.~Ossokine et~al.
\newblock {Multipolar Effective-One-Body Waveforms for Precessing Binary Black
  Holes: Construction and Validation}.
\newblock {\em Phys. Rev. D}, 102(4):044055, 2020.

\bibitem{SEOBNRv3a}
Y.~Pan, A.~Buonanno, A.~Taracchini, L.~E. Kidder, A.~H. Mrou\'e, H.~P.
  Pfeiffer, M.~A. Scheel, and B.~Szil\'agyi.
\newblock {Inspiral-merger-ringdown waveforms of spinning, precessing
  black-hole binaries in the effective-one-body formalism}.
\newblock {\em Phys. Rev. D}, 89(8):084006, 2014.

\bibitem{Pankow2015}
C.~Pankow, P.~Brady, E.~Ochsner, and R.~O'Shaughnessy.
\newblock Novel scheme for rapid parallel parameter estimation of gravitational
  waves from compact binary coalescences.
\newblock {\em Phys. Rev. D}, 92:023002, Jul 2015.

\bibitem{PoissonGW1995}
E.~{Poisson} and C.~M. {Will}.
\newblock {Gravitational waves from inspiraling compact binaries: Parameter
  estimation using second-post-Newtonian waveforms}.
\newblock {\em \prd}, 52:848--855, July 1995.

\bibitem{IMRPhenomXPHMc}
G.~Pratten, C.~Garc{\'{\i} }a-Quir{\'{o}}s, M.~Colleoni, A.~Ramos-Buades,
  H.~Estell{\'{e}}s, M.~Mateu-Lucena, R.~Jaume, M.~Haney, D.~Keitel, J.~E.
  Thompson, and S.~Husa.
\newblock Computationally efficient models for the dominant and subdominant
  harmonic modes of precessing binary black holes.
\newblock {\em Physical Review D}, 103(10), may 2021.

\bibitem{IMRPhenomXPHMa}
G.~Pratten, S.~Husa, C.~Garcia-Quiros, M.~Colleoni, A.~Ramos-Buades,
  H.~Estelles, and R.~Jaume.
\newblock {Setting the cornerstone for a family of models for gravitational
  waves from compact binaries: The dominant harmonic for nonprecessing
  quasicircular black holes}.
\newblock {\em Phys. Rev. D}, 102(6):064001, 2020.

\bibitem{Ramos-Buades-NR}
A.~Ramos-Buades, S.~Husa, G.~Pratten, H.~Estell\'es, C.~Garc\'{\i}a-Quir\'os,
  M.~Mateu-Lucena, M.~Colleoni, and R.~Jaume.
\newblock First survey of spinning eccentric black hole mergers: Numerical
  relativity simulations, hybrid waveforms, and parameter estimation.
\newblock {\em Phys. Rev. D}, 101:083015, Apr 2020.

\bibitem{2022arXiv221106435R}
A.~{Ray}, M.~{Camilo}, J.~{Creighton}, S.~{Ghosh}, and S.~{Morisaki}.
\newblock {Rapid Hierarchical Inference of Neutron Star Equation of State from
  multiple Gravitational Wave Observations of Binary Neutron Star
  Coalescences}.
\newblock {\em arXiv e-prints}, page arXiv:2211.06435, Nov. 2022.

\bibitem{Rinaldi_2021}
S.~Rinaldi and W.~D. Pozzo.
\newblock (h){DPGMM}: a hierarchy of dirichlet process gaussian mixture models
  for the inference of the black hole mass function.
\newblock {\em Monthly Notices of the Royal Astronomical Society},
  509(4):5454--5466, nov 2021.

\bibitem{Sadiq2021}
J.~Sadiq, T.~Dent, and D.~Wysocki.
\newblock Flexible and fast estimation of binary merger population
  distributions with adaptive kde, 2021.

\bibitem{ScargleHistogram2012}
J.~D. {Scargle}, J.~P. {Norris}, B.~{Jackson}, and J.~{Chiang}.
\newblock {Studies in Astronomical Time Series Analysis. VI. Bayesian Block
  Representations}.
\newblock {\em \apj}, 764(2):167, Feb. 2013.

\bibitem{Singer_2016}
L.~P. Singer and L.~R. Price.
\newblock Rapid bayesian position reconstruction for gravitational-wave
  transients.
\newblock {\em Physical Review D}, 93(2), jan 2016.

\bibitem{SEOBNRv4Tb}
J.~Steinhoff, T.~Hinderer, A.~Buonanno, and A.~Taracchini.
\newblock {Dynamical Tides in General Relativity: Effective Action and
  Effective-One-Body Hamiltonian}.
\newblock {\em Phys. Rev. D}, 94(10):104028, 2016.

\bibitem{Talbot2019GWPopulation}
C.~Talbot, R.~Smith, E.~Thrane, and G.~B. Poole.
\newblock Parallelized inference for gravitational-wave astronomy.
\newblock {\em Phys. Rev. D}, 100:043030, Aug 2019.

\bibitem{2020arXiv201201317T}
C.~{Talbot} and E.~{Thrane}.
\newblock {Fast, flexible, and accurate evaluation of gravitational-wave
  Malmquist bias with machine learning}.
\newblock {\em arXiv e-prints}, page arXiv:2012.01317, Dec. 2020.

\bibitem{SEOBNRv3b}
A.~Taracchini et~al.
\newblock {Effective-one-body model for black-hole binaries with generic mass
  ratios and spins}.
\newblock {\em Phys. Rev. D}, 89(6):061502, 2014.

\bibitem{GW190412}
{The LIGO Scientific Collaboration}.
\newblock {GW}190412: Observation of a binary-black-hole coalescence with
  asymmetric masses.
\newblock {\em Physical Review D}, 102(4), aug 2020.

\bibitem{LIGO-O3-O3a-RP}
{The LIGO Scientific Collaboration}, {the Virgo Collaboration}, B.~P. {Abbott},
  R.~{Abbott}, T.~D. {Abbott}, S.~{Abraham}, and et~al.
\newblock {Population properties of compact objects from the second
  LIGO–Virgo Gravitational-Wave Transient Catalog}.
\newblock {\em Available as LIGO-P2000077}, Oct. 2020.

\bibitem{LIGO-GW170817-EOSrank}
{The LIGO Scientific Collaboration}, {the Virgo Collaboration}, B.~P. {Abbott},
  R.~{Abbott}, T.~D. {Abbott}, and {et al}.
\newblock {Model comparison from LIGO-Virgo data on GW170817's binary
  components and consequences for the merger remnant}.
\newblock {\em Classical and Quantum Gravity}, 37(4):045006, Feb. 2020.

\bibitem{Tiwari_2021}
V.~Tiwari.
\newblock {VAMANA}: modeling binary black hole population with minimal
  assumptions.
\newblock {\em Classical and Quantum Gravity}, 38(15):155007, jul 2021.

\bibitem{Varma_2019}
V.~Varma, S.~E. Field, M.~A. Scheel, J.~Blackman, D.~Gerosa, L.~C. Stein, L.~E.
  Kidder, and H.~P. Pfeiffer.
\newblock Surrogate models for precessing binary black hole simulations with
  unequal masses.
\newblock {\em Physical Review Research}, 1(3), oct 2019.

\bibitem{NRSur7dq4}
V.~Varma, S.~E. Field, M.~A. Scheel, J.~Blackman, D.~Gerosa, L.~C. Stein, L.~E.
  Kidder, and H.~P. Pfeiffer.
\newblock Surrogate models for precessing binary black hole simulations with
  unequal masses.
\newblock {\em Physical Review Research}, 1(3), oct 2019.

\bibitem{Veitch2015LALInference}
J.~Veitch, V.~Raymond, B.~Farr, W.~Farr, P.~Graff, S.~Vitale, B.~Aylott,
  K.~Blackburn, N.~Christensen, M.~Coughlin, W.~Del~Pozzo, F.~Feroz, J.~Gair,
  C.-J. Haster, V.~Kalogera, T.~Littenberg, I.~Mandel, R.~O'Shaughnessy,
  M.~Pitkin, C.~Rodriguez, C.~R\"over, T.~Sidery, R.~Smith, M.~Van Der~Sluys,
  A.~Vecchio, W.~Vousden, and L.~Wade.
\newblock Parameter estimation for compact binaries with ground-based
  gravitational-wave observations using the lalinference software library.
\newblock {\em Phys. Rev. D}, 91:042003, Feb 2015.

\bibitem{2020SciPy-NMeth}
P.~Virtanen, R.~Gommers, T.~E. Oliphant, M.~Haberland, T.~Reddy, D.~Cournapeau,
  E.~Burovski, P.~Peterson, W.~Weckesser, J.~Bright, S.~J. {van der Walt},
  M.~Brett, J.~Wilson, K.~J. Millman, N.~Mayorov, A.~R.~J. Nelson, E.~Jones,
  R.~Kern, E.~Larson, C.~J. Carey, {\.I}.~Polat, Y.~Feng, E.~W. Moore,
  J.~{VanderPlas}, D.~Laxalde, J.~Perktold, R.~Cimrman, I.~Henriksen, E.~A.
  Quintero, C.~R. Harris, A.~M. Archibald, A.~H. Ribeiro, F.~Pedregosa, P.~{van
  Mulbregt}, and {SciPy 1.0 Contributors}.
\newblock {{SciPy} 1.0: Fundamental Algorithms for Scientific Computing in
  Python}.
\newblock {\em Nature Methods}, 17:261--272, 2020.

\bibitem{williams2006gaussian}
C.~K. Williams and C.~E. Rasmussen.
\newblock {\em Gaussian processes for machine learning}, volume~2.
\newblock MIT press Cambridge, MA, 2006.

\bibitem{gwastro-RIFT-Update}
J.~Wofford, A.~Yelikar, H.~Gallagher, E.~Champion, D.~Wysocki, V.~Delfavero,
  J.~Lange, C.~Rose, S.~Morisaki, and R.~O'Shaughnessy.
\newblock Expanding rift: Improving performance for gw parameter inference.
\newblock 2022.

\bibitem{2019PhRvD.100h3015W}
K.~W.~K. {Wong} and D.~{Gerosa}.
\newblock {Machine-learning interpolation of population-synthesis simulations
  to interpret gravitational-wave observations: A case study}.
\newblock {\em \prd}, 100(8):083015, Oct. 2019.

\bibitem{2018PhRvD..97d3014W}
D.~{Wysocki}, D.~{Gerosa}, R.~{O'Shaughnessy}, K.~{Belczynski}, W.~{Gladysz},
  E.~{Berti}, M.~{Kesden}, and D.~E. {Holz}.
\newblock {Explaining LIGO's observations via isolated binary evolution with
  natal kicks}.
\newblock {\em \prd}, 97(4):043014, Feb. 2018.

\bibitem{Wysocki2018}
D.~Wysocki, D.~Gerosa, R.~O’Shaughnessy, K.~Belczynski, W.~Gladysz, E.~Berti,
  M.~Kesden, and D.~E. Holz.
\newblock Explaining ligo’s observations via isolated binary evolution with
  natal kicks.
\newblock {\em Physical Review D}, 97(4), Feb 2018.

\bibitem{Wysocki2019}
D.~Wysocki, J.~Lange, and R.~O’Shaughnessy.
\newblock Reconstructing phenomenological distributions of compact binaries via
  gravitational wave observations.
\newblock {\em Physical Review D}, 100(4), Aug 2019.

\bibitem{Wysocki2020}
D.~Wysocki, R.~O'Shaughnessy, L.~Wade, and J.~Lange.
\newblock Inferring the neutron star equation of state simultaneously with the
  population of merging neutron stars, 2020.

\bibitem{Zackay2018}
B.~Zackay, L.~Dai, and T.~Venumadhav.
\newblock Relative binning and fast likelihood evaluation for gravitational
  wave parameter estimation, 2018.

\bibitem{ETK2013}
M.~ZILH{\~{A} }O and F.~LÖFFLER.
\newblock {AN} {INTRODUCTION} {TO} {THE} {EINSTEIN} {TOOLKIT}.
\newblock {\em International Journal of Modern Physics A}, 28(22n23):1340014,
  sep 2013.

\end{thebibliography}

\end{document}